\documentclass{pasj01}
\draft

\begin{document}
\SetRunningHead{Author(s) in page-head}{Running Head}

\title{Reproductions of super-orbital X-ray light curves with the precessing accretion ring model and implications on accretion flows through accretion rings}

\author{Hajime \textsc{Inoue}} %
\affil{Meisei University, 2-1-1 Hodokubo, Hino, Tokyo 191-8506, Japan}
\email{inoue-ha$@$msc.biglobe.ne.jp}


%


\maketitle

\begin{abstract}
X-ray light curves of three X-ray pulsars, SMC X-1, LMC X-4 and Her X-1, 
folded with their respective super-orbital periods, are shown to be well reproduced 
by a model in which X-rays from a compact object towards us are periodically obscured 
by a precessing ring at the outermost part of an accretion disk 
around the central object. 
A situation is considered in which matter from a companion star flows 
into a gravitational field of a compact star carrying a certain amount of 
specific angular momentum 
and first forms a geometrically thick 
ring-tube along the Keplerian circular orbit.
For the model to well fit to the observations, 
it is necessary that the optical depth of the ring-tube for Compton scattering, $\tau \simeq 1 \sim$ 2, the ring matter temperature, $T \simeq 10^{5} \sim 10^{6}$ K 
and the ionization parameter, $\xi \simeq 10^{2}$ erg cm s$^{-1}$ due to X-ray heating from the central X-ray source.  
From simple energetics- and perturbation-arguments, we find that a precession 
of such a ring is rather stable and possible to be excited in the $T$ and $\xi$ ranges.  
The time during which matter accumulates in the ring is estimated to be $\sim 10^{6}$ s, 
and is shown to be comparable 
to the time for an accretion disk to extend from the ring.
It is discussed that in the above $T$ and $\xi $ ranges,
the ring-tube matter could become thermally unstable. 
Then, relatively high density regions in the ring-tube further cools down and tends to shrink to the tube center.  
The flow across the ring circulating flow
should excite turbulent motions, 
and angular momenta of the matter would be effectively transferred across the tube. 
Finally, a steady flow should be established from the companion star through the accretion ring to the accretion disk towards the central compact star.  
\end{abstract}

\section{Introduction}

Several X-ray binaries are known to exhibit super-orbital periods which are 
quasi-periodic on time-scales substantially longer than their orbital periods
(e.g., Priedhorsky \& Holt 1987; Wen et al. 2006; Kotze \& Charles 2012).
A variety of mechanisms have been proposed to account for the long term variabilities. Most of them are briefly summarized by Kotze and Charles (2012) and 
energetics arguments on tidal-force-induced precessions of accretion rings by Inoue (2012) should be added to them.  There have, however, been no trial for any models to 
reproduce the X-ray light curves folded with the super-orbital periods of relevant sources.

In this paper, we try to reproduce observed super-orbital X-ray light 
curves with model light curves.  
We employ a model in which X-rays from a compact object towards us are periodically 
obscured by a precessing ring at the outermost part of an accretion disk 
which was discussed by Inoue (2012).
The target sources are three X-ray pulsars, Her X-1, LMC X-4 and SMC X-1 
which show fairly stable periodic modulations of X-ray intensities on time scales 
of several 10 days in the MAXI public archive data (Matsuoka et al. 2009).

Her X-1 is a low mass X-ray binary with an orbital period of 1.7 d  
(for its binary parameter, see e.g. Leahy \& Abdallah 2014).
It has a 35 day period of X-ray on-off, which was discovered with UHURU 
(Giacconi et al. 1973).  The 35-day cycle sometimes becomes unclear during which 
the X-ray intensity is very low (called as anomalous low state, e.g. Parmar et al. 1985).  
LMC X-4 and SMC X-1 are massive X-ray binaries with orbital periods of 1.4 d and 
3.9 d respectively (for their binary parameters, see e.g. Falanga et al. 2015).
LMC X-4 exhibits a 30 day period of X-ray modulation which was 
found with HEAO-1 by Lang et al. (1981).  The 30 day cycle is seen steadily 
(Clarkson et al. 2003b; Kotze \& Charles 2012) and no significant period excursion has been reported.
SMC X-1 shows a $\sim$ 60 day period of X-ray modulation, which was first suggested 
from HEAO-1 observations by Gruber and Rothschild (1984) and confirmed with RXTE 
and CGRO observations by Wojdowski et al. (1998).
This period is, however, known to sometimes change from $\sim$ 60 d 
to $\sim$ 40 d and return to the former $\sim$ 60 d (Clarkson et al. 2003a; Hu et al. 2011).

For these sources, several observational evidences are pointed out to be consistent 
with a picture that 
the high-low cycle in the X-ray flux is the result of a quasi-periodic occultation 
of the central X-ray source by a precessing tilted accretion disk, 
by e.g. Jones and Forman (1976); Gerend and Boynton (1976) for Her X-1, 
Heemskerk and van Paradijs (1989) for LMC X-4, and Wojdowski et al. (1998) for SMC X-1.

The mechanism of the precessing tilted accretion disk has been discussed by 
several authors.
After Katz (1973), Levine and Jernigan (1982) considered precessing motion of a ring 
of given radius in an accretion disk which is tilted
with respect to the binary orbital plane.   It is naturally understood 
that tidal force from the companion star exerts a torque on the tilted ring and it causes the precessing motion.
Questions are, however, how one precession rate corresponding to a particular ring-radius in the disk is selected and how the disk holds the tilt against viscous damping.
Radiation-induced warping and tilting of accretion disks was, then, proposed by 
Pringle (1996) and developed by Wijers and Pringle (1999); Ogilvie and Dubus (2001).
Based on an instability of the disk through the radiation reaction force 
by X-rays from the central source, the disk is shown to steadily tilt and precess, 
although parameter ranges for the steadily precessing disk are discussed to be narrow.

Inoue (2012) proposed a different idea.  He paid attention to a ring 
at the outermost part of an accretion disk which matter from the companion star 
should form along a Keplerian circular orbit determined by the average specific angular 
momentum of the inflowing matter around the compact star.  
He examined energetics of a precession of the ring under a tidal force from the 
companion star when the ring rotation axis tilts from the binary rotation axis.
It was found that the energy minimum exists at a certain, non-zero tilt angle 
when the ring has sufficient thermal energy.
Here, we add simple perturbation arguments on excitation of the precession 
in appendix 1.  These energetics and perturbation arguments indicate 
that the precession of the ring should be realized as the most stable state.
This precessing accretion ring model can well explain the two key questions 
on the precessing accretion disk models, 
how a particular precession period is selected and how a precession is kept steady,  
by the firm presence of the ring at the outermost part of the accretion disk and 
the stability of the precession respectively.

Hereafter, we fit X-ray profiles expected from the model, that 
X-rays from compact stars towards us are periodically obscured by the precessing ring,  
to X-ray light curves folded with the respective super-orbital periods
of three X-ray pulsars, Her X-1, SMC X-1 and LMC X-4 observed with MAXI.  
The methods and results are shown in section 2.  Discussions on the precessing rings 
from the best fit parameters and more general discussions on accretion flows 
through accretion rings are given in sections 3 and 4 respectively.
 
\section{Analysis and Results}

We analyze ''orbit" data of three X-ray pulsars, Her X-1, SMC X-1 and LMC X-4, obtained 
from the MAXI public data.

\subsection{Search of super-orbital periods and folded light curves}
We searched super-orbital periods of the three pulsars by the epoch-folding analysis (Priedhorsky \& Terrell, 1983), with looking for a period at which the $\chi^2$ value of the folded light curve against the constant average flux becomes maximum.
Initially, we did it for the whole data (Aug, 2009 $\sim$ Feb. 2016) from the MAXI archive 
and checked whether the periodicity is steadily seen in the entire light curve, for each of the three sources.\\

\begin{figure}
  \begin{center}
    \FigureFile(100mm,100mm){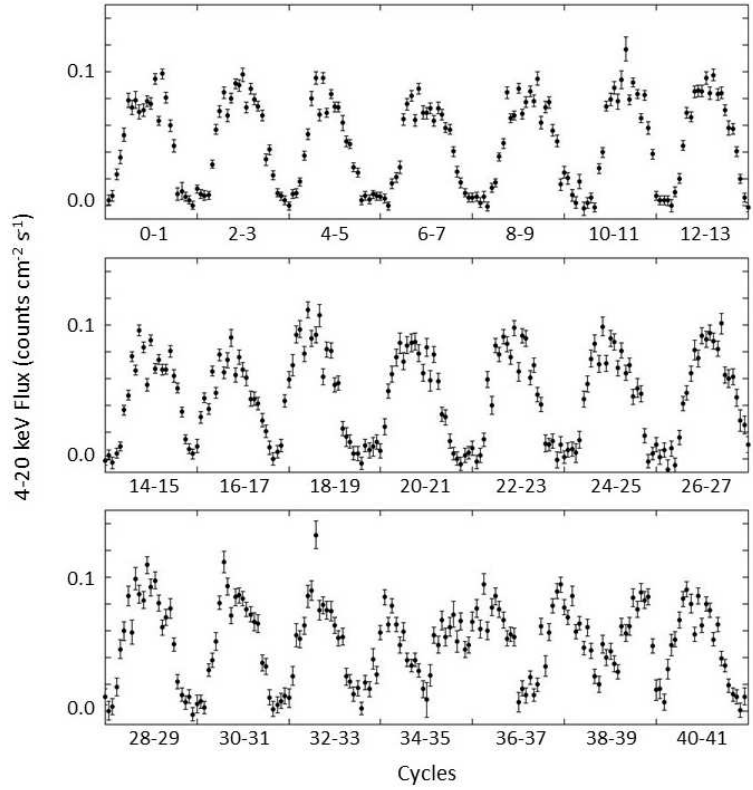}
  \end{center}
  \caption{4 - 20 keV light curves folded every two cycles with the 55.73 day super-orbital period of SMC X-1.}
\label{SMCX-1_LC}
\end{figure}

For SMC X-1, the most likely super-orbital period was first found to be 55.6 days
but an apparent period shift was seen after the cycle number get to $\sim$ 32.  
Hence, we have limited the data with the cycle of less than 32 and have obtained the revised most likely period 
to be 55.73 days through the epoch-folding analysis of the limited data.
Figure \ref{SMCX-1_LC} shows a 4-20 keV X-ray light curve folded every two cycles with the 55.73 
day period.  You can clearly see that the super-orbital periodicity becomes unclear when the cycle  exceeds $\sim$ 32.

This period change seems to be similar to drifts of the super-orbital period 
from the normal 50 $\sim$ 60 days to $\sim$ 45 days reported by Clarkson et al. (2003a) and Hu et al. (2011).  
Although this is an interesting event, we do not go into the detail here.

\begin{figure}
  \begin{center}
    \FigureFile(80mm,80mm){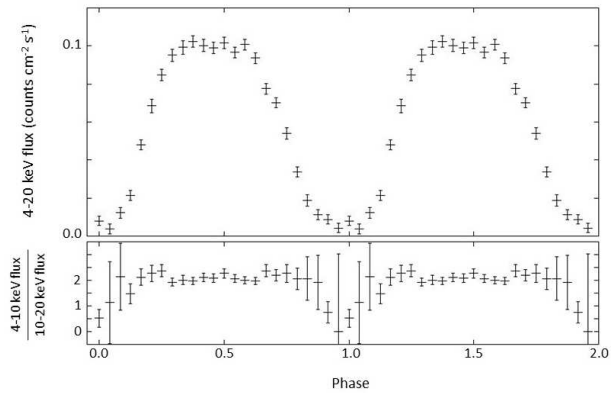}
  \end{center}
  \caption{4-20 keV super-orbital light curve of SMC X-1 
folded over 16 selected cycles (upper panel) 
and phase-dependence of the softness ratio of 4-10 keV flux to 10-20 keV flux(lower panel). 
Two cycles are plotted.  Vertical lines represent statistical errors propagated from 
the original errors in the MAXI archive data.}
\label{SMCX-1_foldedLC}
\end{figure}

Ever if we limit the cycle number less than 32, there yet remains small but significant 
phase jitters as seen in figure  \ref{SMCX-1_LC}.  
Thus, we have further limited the cycles to get the super-orbital light curve 
as appropriate as possible and the selected eight two-cycle sets are 0-1, 2-3, 4-5, 14-15, 22-23, 24-25, 28-29 and 30-31.   

The finally obtained the super-orbital light curve of SMC X-1 is presented in figure \ref{SMCX-1_foldedLC}. 
The softness ratios of the 4-10 keV flux to the 10-20 keV flux are also plotted against the super-orbital phases.  Although the ratio tends to decrease around the flux minimum, the spectral softening is not statistically significant.

It should be noted here is that the data used in figures \ref{SMCX-1_LC} and 
\ref{SMCX-1_foldedLC} exclude 
data during the orbital eclipses.  We made a light curve folded with the orbital period and selected a range of the orbital phase free from X-ray eclipses.  
The same processes have been taken for LMC X-4 and Her X-1.

\begin{figure}
  \begin{center}
    \FigureFile(100mm,100mm){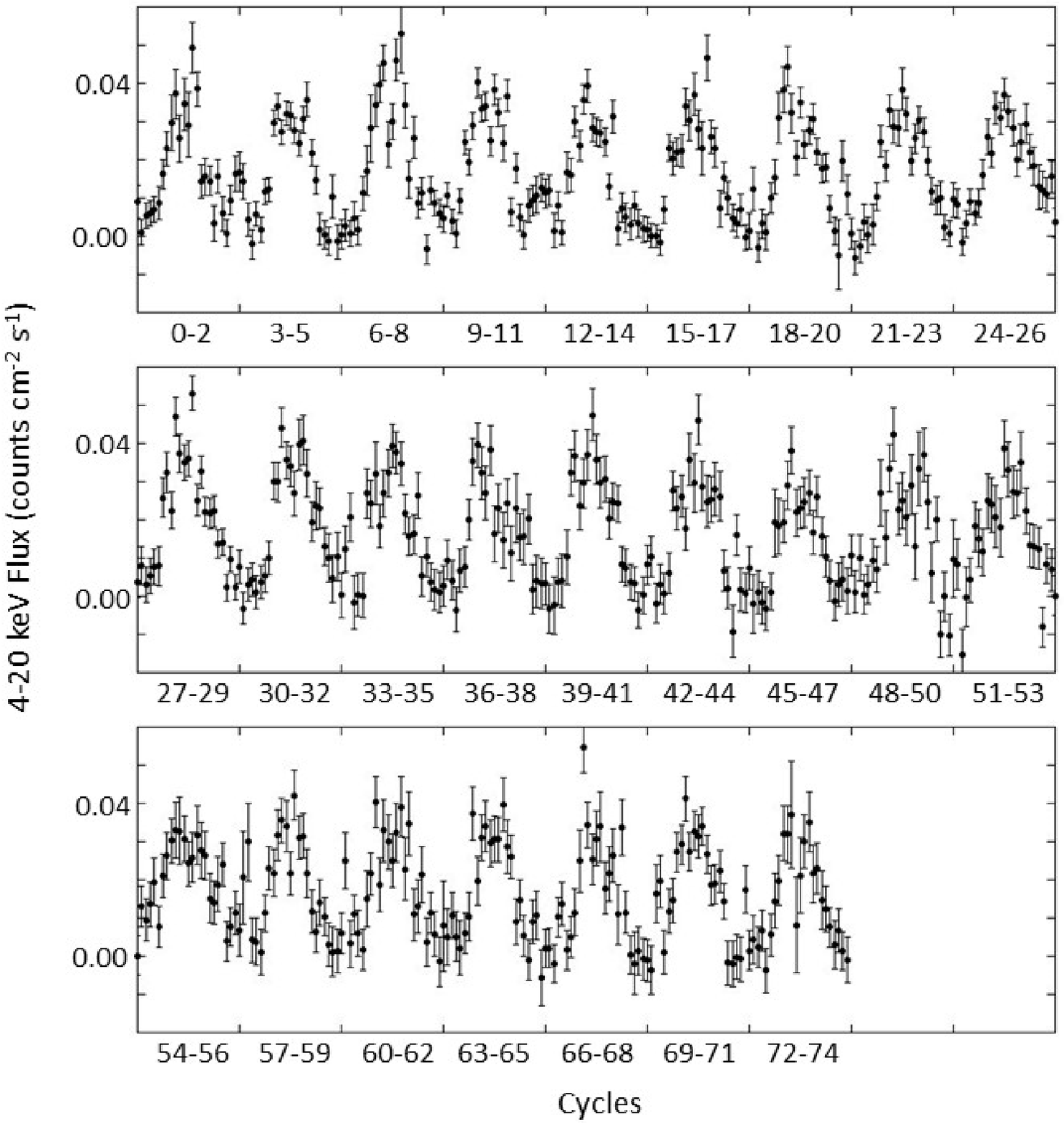}
  \end{center}
  \caption{4 - 20 keV light curves folded every three cycles with the 30.39 days super-orbital period of LMC X-4.}\label{LMCX-4_LC}
\end{figure}

\begin{figure}
  \begin{center}
    \FigureFile(80mm,80mm){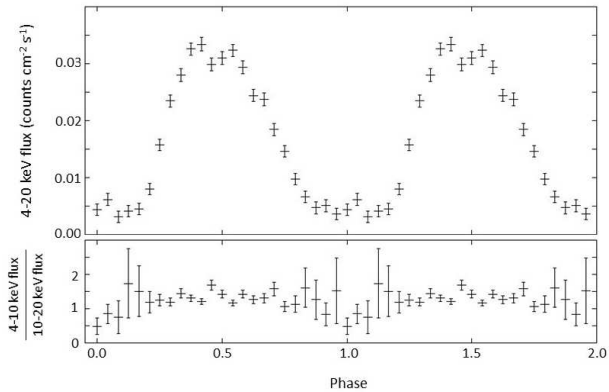}
  \end{center}
  \caption{4-20 keV super-orbital light curve of LMC X-4 
folded over 75 cycles (upper) and the softness ratio variation (lower).}\label{LMCX-4_foldedLC}
\end{figure}

For LMC X-4, the most likely super-orbital period was 30.39 days.
Figure \ref{LMCX-4_LC} shows a 4-20 keV X-ray light curve folded every three cycles with that period. 
Folding every three cycles was needed for this source to have a complete data set 
for the error estimation as in appendix 2.
As seen from figure \ref{LMCX-4_LC}, this source shows no large period shift nor large amplitude change in terms of the super-orbital periodicity.  Thus, we have folded all 75 cycles and the folded light curve is given together with the softness ratio variation 
in figure \ref{LMCX-4_foldedLC}.

\begin{figure}
  \begin{center}
    \FigureFile(100mm,100mm){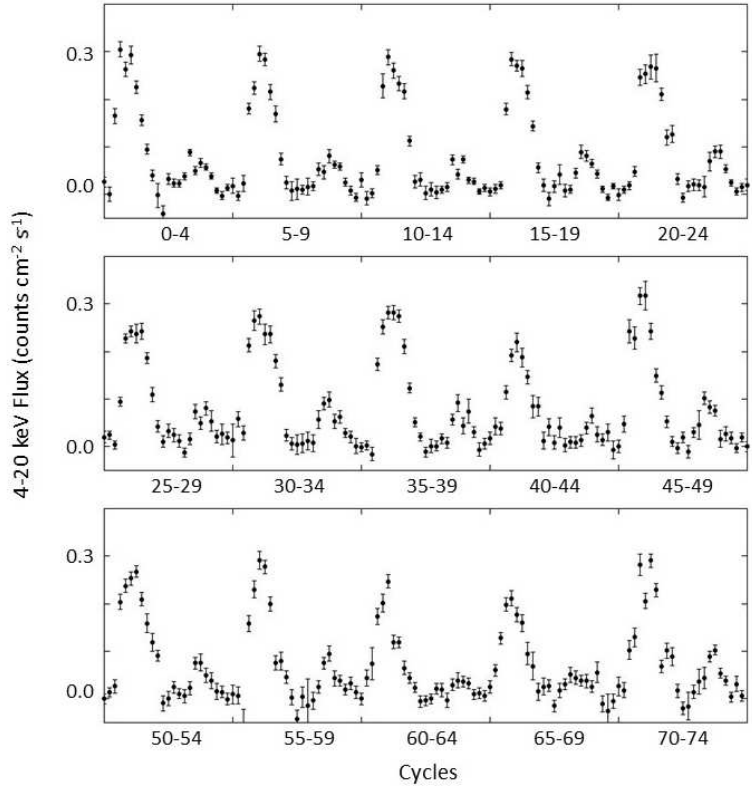}
  \end{center}
  \caption{4 - 20 keV light curves folded every five cycles with the 34.84 day super-orbital period of Her X-1.}\label{HerX-1_LC}
\end{figure}

\begin{figure}
  \begin{center}
    \FigureFile(80mm,80mm){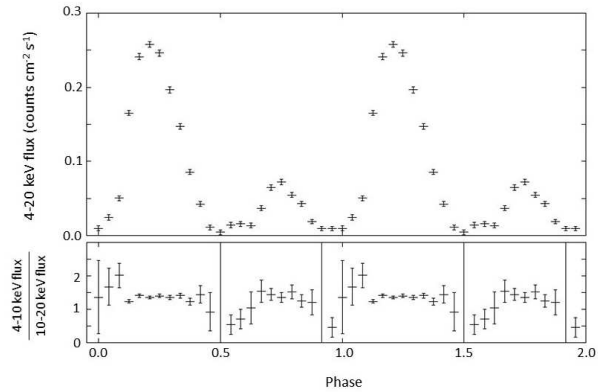}
  \end{center}
  \caption{4-20 keV super-orbital light curve of Her X-1 
folded over 75 cycles (upper) and the softness ratio variation (lower).}\label{HerX-1_foldedLC}
\end{figure}

For Her X-1, the most likely super-orbital period was 34.84 days.
Figure \ref{HerX-1_LC} shows a 4-20 keV X-ray light curve folded every five cycles with that period. 
For the error estimation, we needed folding every five cycles for this source.
The 4-20 keV super-orbital light curves folded every five cycles are plotted in figure \ref{HerX-1_foldedLC}.
Although the amplitude fluctuations and the phase modulations are seen, they are not so large as to limit the cycle-sets to obtain the average profile. 
Thus, we have made the super-orbital light curve from all the cycle sets, and the result is presented in figure \ref{HerX-1_foldedLC}.  The softness ratio variation is also plotted there.\\

\subsection{Model fits}

Next, the folded light curves of the three X-ray pulsars have respectively been fit to the  precessing ring model
that X-rays from a compact object towards us are periodically obscured by a precessing ring at the outermost part of an accretion disk around the central object.

For the errors in the folded light curves used in the model fits, 
a simple propagations of the original errors attached in the MAXI archive data would be insufficient, since X-ray flux from the source 
have a fairly large time variation on the time scale of the phase-bin width.
Furthermore, small phase-jitters exist and cause some amount of flux excursion at phases with large flux gradients.
Thus, we have adopted such error estimations as in appendix 2.

\begin{figure}
  \begin{center}
    \FigureFile(80mm,80mm){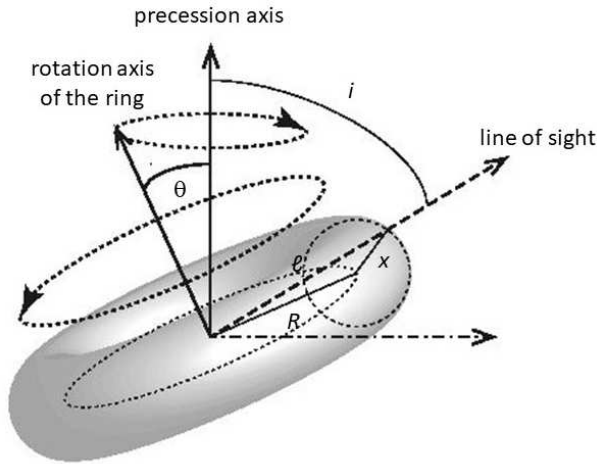}
  \end{center}
  \caption{Configuration and geometrical parameters of the precessing ring model.  
The density distribution of matter in the direction of $x$ is assumed to be as given 
in equation (\ref{eqn:n_def}) with a scale height parameter, $x_{0}$, defined in equation (\ref{eqn:a_def}).
}\label{PrecessingRing}
\end{figure}

The model configuration is schematically depicted in figure \ref{PrecessingRing}.  
A ring is precessing with a tilt angle, $\theta$, from the equatorial plane of a binary 
system around a precession axis normal to the binary equatorial plane.  
Matter in the ring is assumed to have the same angular momentum per mass, 
and the ring radius, $R$, around the central neutron star is determined 
by a balance between the centrifugal force from the circular motion 
with the given angular momentum and the gravitational force from the neutron star.  
Since the ring matter has thermal energy, the ring has a thickness in directions 
perpendicular to the Keplerian circular orbit and forms a tube like structure 
along the circular orbit.  The density distribution of the matter in a cross section of the ring-tube is given in equation (\ref{eqn:n_def}) under an isothermal 
approximation and it has two parameters, the gaseous number density 
at the tube center, $n_{0}$ and the scale height factor, $x_{0}$ defined 
in equation (\ref{eqn:a_def}).
Premises of this model and the structure of the ring-tube are described in the introductory part of appendix 1.   

X-rays from the neutron star vicinity are assumed to be scattered through 
electron scatterings by the ring tube matter and the degree of the obscuration is calculated by an equation as $\exp(-\tau)$ where $\tau$ is the optical depth for 
electron scattering along the line of sight, given as
\begin{equation}
\tau = \sigma_{\rm{T}} \int n dl.
\label{eqn:tau_def}
\end{equation}
$\sigma_{\rm{T}}$ is the cross section of the Thomson scattering and the integration is 
done along the line of sight from the neutron star to the observer.
As the ring axis precesses, the column density along the line of sight changes and 
the obscuration of the observed X-rays periodically varies.

To the model fits, the simple model has first been adopted, in which a rotationally uniform ring-tube steadily 
preccsses with a tilt angle.
Free parameters are seven, 
an inclination angle of the line of sight against the precession axis, $i$,
a tilt angle of the ring axis against the precession axis, $\theta$,
a relative thickness of the ring tube, $x_{0}/R$,
a typical electron column density of the ring tube, $N_{\rm e} = n_{0} x_{0}$, 
an intrinsic X-ray flux suffering from periodic obscuration by the precessing ring, $F_{0}$,
a constant X-ray flux scattered by the ring matter, $F_{\rm C}$, 
and a precession phase at the data start time.

The dependences of the $\chi^{2}$ minimum values on $i$ are found to be very weak 
and we cannot constrain the 90\% confidence ranges of $i$ for either of the three pulsars.
Hence, we have fixed values of $i$ at typical values from the literature on  
binary parameters of the three sources.
For SMC X-1 and LMC X-4, 1 $\sigma$ level uncertainty ranges are obtained 
by Falanga et al. (2015), to be 62$^{\circ}$ $\pm$ 2$^{\circ}$ and 59.3$^{\circ}$ $\pm$ 
0.9$^{\circ}$ respectively.
Then, 62$^{\circ}$ and 59$^{\circ}$ have been used as the fixed $i$ values respectively for SMC X-1 and LMC X-4.
For Her X-1, a fairly large uncertainty range is given by Leahy and Abdallah (2014) 
to be  80$^{\circ} \sim 90^{\circ}$.
Thus, three cases of 80$^{\circ}$, 83$^{\circ}$ and 86$^{\circ}$ have been adopted.

\begin{figure}
  \begin{center}
    \FigureFile(80mm,80mm){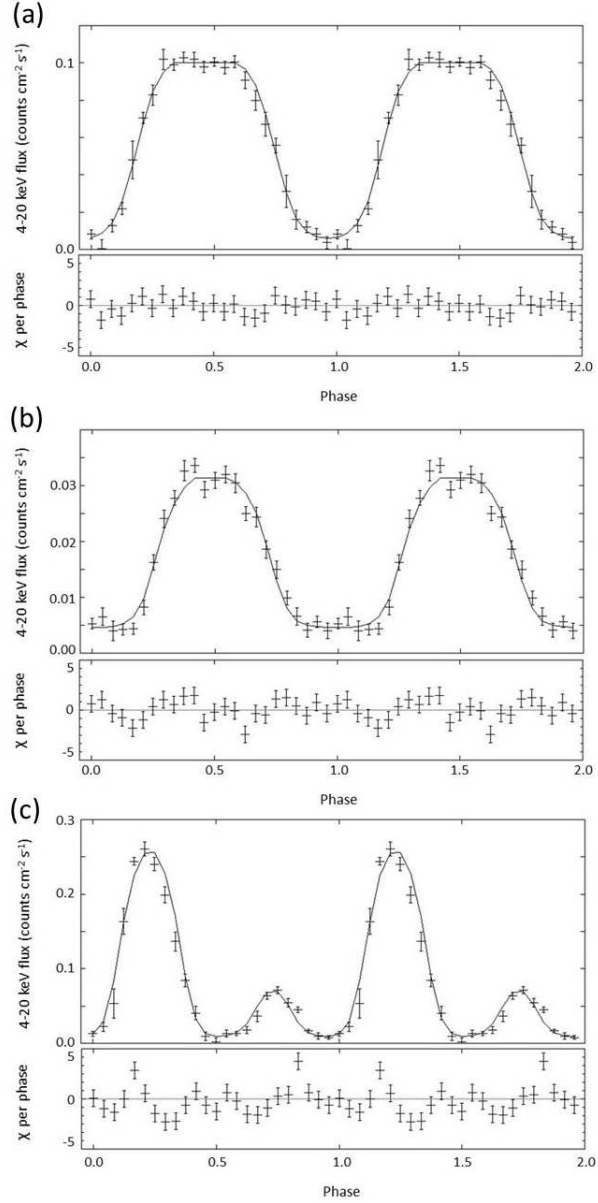}
  \end{center}
  \caption{Plots of the best fit model curve and the observed light curve 
(upper panel) and the distribution of $\chi$ = (model - data) / error at each phase-bin (lower panel) for SMC X-1 (a), for LMC X-4 (b) and for Her X-1 (c).
The simple model is applied and the best fit curves are shown with respective thin lines.  The observed light curves are given with crosses in which vertical lines represent errors estimated as in appendix 2.  
}\label{SimpleModel_BestFitCurves}
\end{figure}

The best fit light curves with the simple model for the three sources 
are plotted together with the observed 
light curves in figure \ref{SimpleModel_BestFitCurves}.  For Her X-1, the result of $i = 83^{\circ}$ is shown as the representative of the three $i$ cases.
The obtained $\chi^{2}_{\nu}$ values are 
0.975 for SMC X-1 (DOF = 19), where the constant X-ray flux has been unnecessary, 
1.941 for LMC X-4 (for DOF = 18), and almost 4 for every three $i$ cases of Her X-1 (DOF = 18). 

The simple precessing ring model is acceptable to the super-orbital light  
curve of SMC X-1.
It is, however, not acceptable to those of LMC X-4 nor Her X-1.
As seen from figures \ref{SimpleModel_BestFitCurves}-(b) and (c), the super-orbital light curves of LMC X-4 and Her X-1 
exhibit asymmetric slopes in the rising phase and the decaying phase.  The simple precessing ring model cannot reproduce 
such an asymmetric shape and it causes the large reduced $\chi^{2}$ value.
Note that even in the case of SMC X-1, the sharp rise and the slow decay seem to exist, although they are not statistically significant.

To improve this, we have phenomenologically introduced modified models 
with a wavy modulation on the ring orbit and the ring-tube thickness respectively.

As for the ring orbit modulation, we have modeled that the orbit of the ring-tube center 
deviates to a direction perpendicular 
to the original orbital plane with angle as viewed from the orbital center, 
$\delta$, as
\begin{equation}
\delta = \delta_{0} \sin 2 (\phi - \phi_{0}) ,
\label{eqn:DeltaMod}
\end{equation}
where $\delta_{0}$ is the amplitude of the wavy deviation, 
$\phi$ is the rotational angle of the precession axis from the direction 
when the vernal equinox of the precessing orbit is in a meridian 
including the line of sight and $\phi_{0}$ is the phase shift angle of the wavy deviation.
The model has become acceptable for LMC X-4 but has not for Her X-1.  
The $\chi_{\nu}^{2}$ value is 0.873 (DOF = 17) for LMC X-4, where the constant X-ray flux, $F_{\rm C}$, is unnecessary.  For Her X-1, those are 2.07 when $i = 80^{\circ}$, 
1.98 when $i = 83^{\circ}$, and 1.82 when $i = 86^{\circ}$, where DOF = 16 including 
$F_{\rm C}$ as a free parameter.

As for the ring-tube thickness modulation, on the other hand, the scale height parameter of the ring tube, $x_{0}$, has been thought to deviate from the average with a fraction given by the same formula as in equation (\ref{eqn:DeltaMod}).  
In this case, the tube-center density, $n_{0}$, should also have a modulation 
in proportion to $x_{0}^{-2}$, since the mass flow continuity along the ring-tube gives 
a relation as $n_{0} x_{0}^{2} =$ constant with the constant flow velocity.
This modified model has given acceptable fits both for LMC X-4 and Her X-1. 
The  $\chi_{\nu}^{2}$ value is  1.242 for LMC X-4 (DOF = 16), where the constant X-ray flux, $F_{\rm C}$, is necessary in this case.  
For Her X-1, those are 0.971 when $i = 80^{\circ}$, 
1.02 when $i = 83^{\circ}$, and 1.26 when $i = 86^{\circ}$, where DOF = 16 including 
$F_{\rm C}$.

From these results, we have employed the ring-tube thickness 
modulation model as the most appropriate model to reproduce 
the observed super-orbital light curves.

\begin{figure}
  \begin{center}
    \FigureFile(80mm,80mm){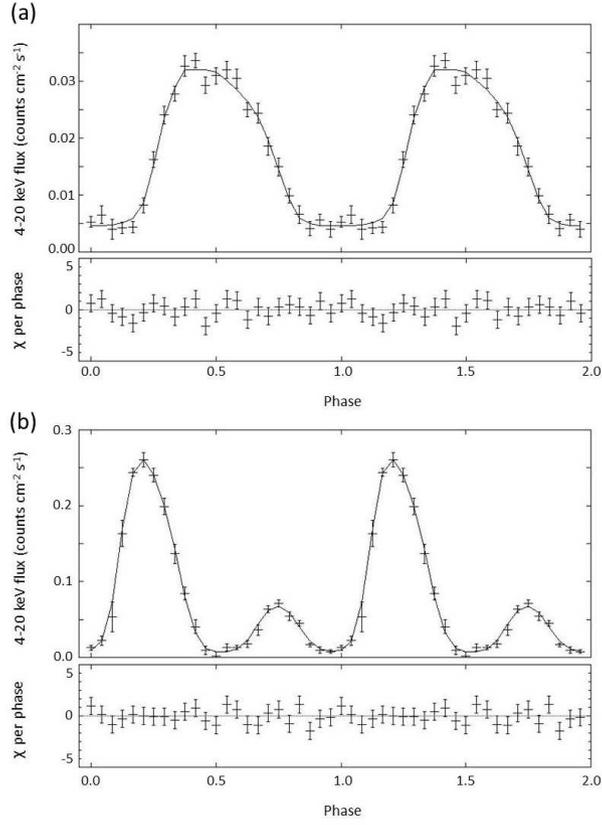}
  \end{center}
  \caption{Plots of the best fit model curve and the observed light curve 
(upper panel) and the distribution of $\chi$ at each phase-bin (lower panel) 
for LMC X-4 (a) and for Her X-1 (b).
The ring-tube thickness modulation model is applied and the best fit curves are shown with respective thin lines.  The observed light curves are given with crosses in which vertical lines represent errors estimated as in appendix 2.  }\label{RTM_Model_BestFitCurves}
\end{figure}

The best fit curves of the ring-tube thickness modulation model for LMC X-4 and Her X-1 are shown in figure   \ref{RTM_Model_BestFitCurves}.

Best fit parameters and their uncertainty ranges obtained from the model fits
are tabulated in table \ref{BestFitValues}. 

\begin{table}
  \caption{Best fit values and 90\% confidence ranges of model parameters}
  \begin{center}
    \begin{tabular}{c|cccccccc}
      \hline
source & $i$ (fixed) & $\theta$ & $N_{\rm e}$ & $x_{0}/R$ & $\delta_{0}$ & $\phi_{0}/2\pi$  & $F_{0}$ & $F_{\rm C}$ \\
     \hline
      SMC X-1 & 62 & 18 $^{+6} _{-6}$ & 2.4 $^{+4.8} _{-0.8}$ & 0.20 $^{+0.02} _{-0.02}$ & - & - & 0.100 $^{+0.001} _{-0.002}$ & -  \\ \hline 
LMC X-4 & 59 & 23 $^{+7} _{-12}$ & 3.4 $^{+4.7} _{-1.3}$ & 0.26 $^{+0.05} _{-0.04}$ & 0.10 $^{+0.03} _{-0.05}$ & -0.03 $^{+0.01} _{-0.07}$ & 0.028 $^{+0.025} _{-0.002}$ & 0.0044 $^{+0.0004}_{-0.0011}$ \\ \hline  & 80 & 40 $^{+12} _{-4}$ & 3.7 $^{+1.8} _{-0.5}$ & 0.44 $^{+0.03} _{-0.03}$ & 0.24 $^{+0.10} _{-0.07}$ & -0.08 $^{+0.02} _{-0.02}$ & 1.2 $^{+0.5} _{-0.5}$ & 0.008 $^{+0.002} _{-0.002}$ \\
  Her X-1 & 83 & 30 $^{+7} _{-4}$ & 3.5 $^{+1.1} _{-0.3}$ & 0.37 $^{+0.03} _{-0.06}$ & 0.19 $^{+0.11} _{-0.04}$ & -0.08 $^{+0.02} _{-0.01}$  & 1.5 $^{+0.9} _{-0.9}$ & 0.006 $^{+0.004} _{-0.002}$ \\  & 86 & 26 $^{+2} _{-8}$  & 5.2 $^{+0.4} _{-2.3}$ & 0.26 $^{+0.01} _{-0.08}$ & 0.26 $^{+0.03} _{-0.14}$ & -0.10 $^{+0.07} _{-0.01} $ & 2.3 $^{+1.3} _{-1.9}$  & 0.010 $^{+0.002} _{-0.006}$ \\
      \hline
 (unit) & degree & degree & 10$^{24}$ cm$^{-2}$ &  & & & counts cm$^{-2}$ s$^{-1}$ & counts cm$^{-2}$ s$^{-1}$
  \\  \hline
    \end{tabular}
  \end{center}
  \label{BestFitValues}
\end{table}

\section{Discussion on the precession ring model}

\subsection{The ring-tube thickness modulation}

We have tried to reproduce X-ray light curves of three X-ray pulsars, 
SMC X-1, LMC X-4 and Her X-1, folded with their respective super-orbital periods, 
by the precessing accretion ring model. 

\begin{figure}
  \begin{center}
    \FigureFile(120mm,80mm){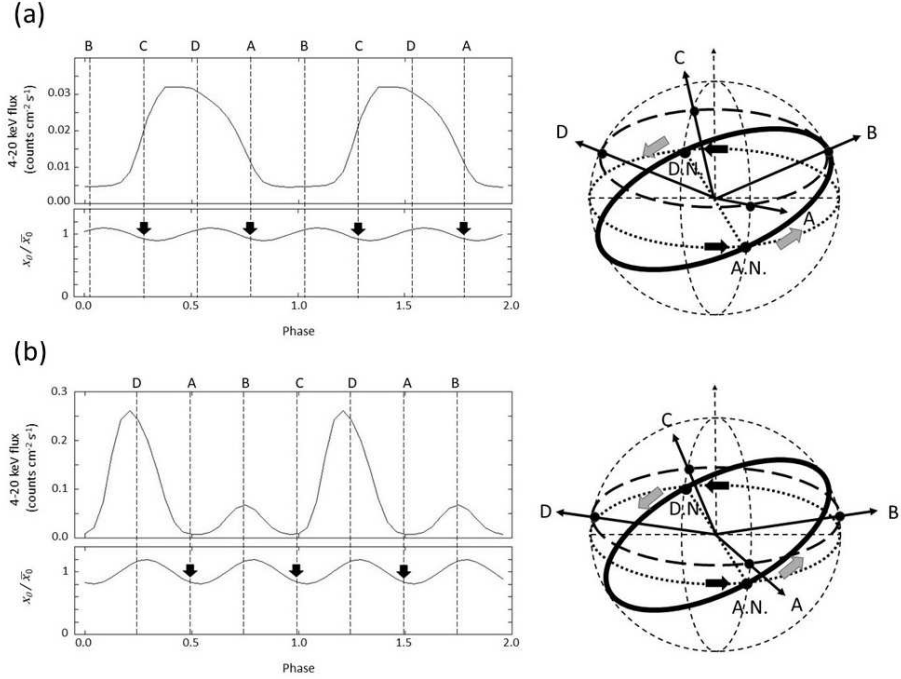}
  \end{center}
  \caption{Best fit model light curves  and modulations of the ring-tube thickness 
(left panels) and the configurations of the ring and the line of sights (right panels) 
for LMC X-4 (a) and Her X-1 (b).  For the details, see the text.  
}\label{ModelConfig}
\end{figure}

Although the model light curve with the simple uniform ring is well fit to 
the super-orbital light curve of SMC X-1, 
a wavy modulation on the ring-tube thickness is necessary 
to get acceptable fits to light curves 
of LMC X-4 and Her X-1.
The ring-tube thickness is represented with the scale-height parameter, $x_{0}$, 
in the model and 
the best fit modulations of $x_{0}$ around the average $\bar{x_{0}}$ are shown, 
together with the best fit model light curves, 
respectively for LMC X-4 and Her X-1 in left-side panels of figure \ref{ModelConfig}.

The schematic ring configurations are also depicted in the right side panels of figure \ref{ModelConfig}.
A thick elliptic line represents the circular orbit of the ring-tube center, and it tilts 
from the orbital plane of the binary which is expressed with a dotted elliptic line, 
in each of the right panels.  Matter flowing from the companion star tends to circulate 
around the compact star and should collide and merge with the ring-tube.
Since the ring-tube tilts, the impacts should take place around 
the ascending and descending nodes, 
which are indicated with A.N. and D.N. respectively in the panels.
The inflow directions towards the node points are assumed to be as with 
black arrows and the circulating direction of the ring matter is indicated with gray arrows.
The ring rotational axis should circulate around the binary rotational axis with 
the ring precession period 
in the anti-direction to the ring circulating direction in an observation frame 
in which the line of sight is fixed.
In the right panels of figure \ref{ModelConfig}, however, the ring orientation is rather fixed.
In such a ring-rest frame, the line of sight between the central X-ray source and 
the observer should rotate around the binary rotational axis in the same direction 
as the ring matter circulation. 
The small circle along which the line of sight rotates is shown with a thick dashed line 
in each of the two right panels, by taking account of the respective inclination angle.
Then, four arrows from the center, crossing the small circle,  with letters, A to D, 
represent the line of sight direction which is in the meridian plane including the ascending node, the upper solstice, the descending node and the lower solstice, respectively.
The four epochs are indicated with thin dashed lines with labels, A to D, in the left-side panels.  We can see the phase of the wavy modulation in relation to the four epochs 
of the line of sight direction.

We see from figure \ref{ModelConfig}-(a) that the line of sight roughly lies on 
the ring plane and the central X-ray source is the most largely obscured 
by the ring matter around the epoch of B in case of LMC X-4.
This is consistent with the argument based on high-spectral-resolution 
X-ray observations of LMC X-4 that the precessing disk is viewed roughly edge-on 
in the X-ray low state, by Neilsen et al. (2009).
On the other hand, figure \ref{ModelConfig}-(b) shows that the line of sight crosses 
the ring plane twice near the epochs A and C and that the largest X-ray obscurations 
appear around these two epochs, in case of Her X-1.  

A possible origin of the wavy modulation could be impacts of the inflow stream from 
the companion star to the precessing ring.
The inflowing matter should hit the ring matter around the ascending and the descending nodes and tend to compress the ring tube.
Since the inflowing matter should soon merge with the circular motion of 
the ring tube matter, however, the ring-tube compression should weaken 
after the inflowing matter flows 
some distance from the impact point and the pressure force of the ring-tube  strengthened by the compression should push back the ring-tube matter there.

In the left panels in figure \ref{ModelConfig}, the timings of the inflow-impact are shown in black arrows 
at the epochs of A and C.  From these panels, we can see that the most compressed 
points in the $x_{0}$ modulation appear slightly after the inflow impacts in both cases of (a) and (b). 
This is consistent with the above considerations.

The ring-tube thickness modulation introduced in the model fits could reasonably be explained by considering the impacts of the inflowing matter to the circulating ring-tube matter.\\

If we accept the precessing ring model as the origin of the super-orbital periods 
of the three X-ray pulsars, we can discuss the physical properties 
of the precessing accretion ring as follows.

\subsection{Geometrical properties}
Binary parameters of the three source are given in table \ref{BinaryParameters}, which are 
from Falanga et al. (2015) for SMC X-1 and LMC X-4, and 
from Leahy and Abdallah (2014) for Her X-1.  
83$^{\circ}$ is used as a representative value of the inclination angle of Her X-1. 

\begin{table}
  \caption{Binary parameters of the three sources}
  \begin{center}
    \begin{tabular}{c|ccc}
      \hline
parameter & SMC X-1 & LMC X-4 & Her X-1 \\
     \hline
      binary period, $P_{\rm B}$ & 3.9 d & 1.4 d & 1.7 d  \\ 
inclination angle, $i$ & 62$^{\circ}$ & 59$^{\circ}$ & 83$^{\circ}$  \\ 
mass of the compact star, $M_{\rm X}$  & 1.2 $M_{\odot}$  & 1.6 $M_{\odot}$ & 1.5 $M_{\odot}$ \\
  mass of the companion star, $M_{\rm C}$ & 18 $M_{\odot}$ & 18 $M_{\odot}$ & 2.3 $M_{\odot}$ \\
separation of the two star, $a$  & $1.9 \times 10^{12}$ cm & $9.9 \times 10^{11}$ cm & $6.6 \times 10^{11}$ cm \\
      \hline
    \end{tabular}
  \end{center}
  \label{BinaryParameters}
\end{table}

The ratio between the radius of the precessing ring, $R$ and the binary separation, $a$ 
is given by Inoue (2012) as
\begin{equation}
\frac{R}{a} = \Bigl(2 \frac{(1 + q)^{1/2}}{q} \frac{P_{\rm B}}{P_{\rm P}} \frac{1}{\cos \theta} \Bigr)^{2/3},
\label{eqn:RperD}
\end{equation}
where $q = M_{\rm C} / M_{\rm X}$ and $P_{\rm P}$ is the precession period.
From this equation, the ring radius, $R$ is calculated and the ring tube radius, $x_{0}$, 
is obtained from the best fit value of $x_{0}/R$.  The results are tabulated in table \ref{GeometricalParameters}.

\begin{table}
  \caption{Geometrical parameters of the precessing ring}
  \begin{center}
    \begin{tabular}{c|ccc}
      \hline
parameter & SMC X-1 & LMC X-4 & Her X-1 \\
     \hline
$q$ & 14.9 & 11.6 & 1.55 \\
$P_{\rm P}$ & 56 d & 30 d & 35 d \\
      $R / a$ & $1.1 \times 10^{-1}$ & $9.3 \times 10^{-2}$ & $2.2 \times 10^{-1}$  \\ 
$R$ & $2.1 \times 10^{11}$ cm & $9.2 \times 10^{10}$ cm & $1.4 \times 10^{11}$ cm  \\ 
$x_{0}/R$  & $2.0 \times 10^{-1}$  & $2.6 \times 10^{-1}$ & $3.7 \times 10^{-1}$ \\
$x_{0}$ & $4.3 \times 10^{10}$ cm & $2.4 \times 10^{10}$ cm & $5.3 \times 10^{10}$ cm \\
      \hline
    \end{tabular}
  \end{center}
  \label{GeometricalParameters}
\end{table}

The radius of the Roche lobe around the compact star, $R_{\rm RL}$, can be calculated 
as a ratio to the separation, $a$, from the approximate form by Eggleton (1983) as 
\begin{equation}
\frac{R_{\rm RL}}{a} = \frac{0.49}{0.6 + q^{2/3} \ln (1 + q^{-1/3})}.
\label{eqn:RperRRL}
\end{equation}
The values of $R_{\rm RL} / a$ are 0.18, 0.20 and 0.34 for SMC X-1, LMC X-4 and Her X-1 respectively.  By comparing these values with the values of $R/a$ in table \ref{GeometricalParameters}, we see that the ring radii, $R$, are all reasonably smaller than the Roche lobe radii, $R_{\rm RL}$.

\subsection{Gaseous properties of the ring matter}

The temperature, $T$, of the gaseous matter in the ring can be calculated 
from equation (\ref{eqn:a_def}) by using the values of $x_{0}$ and $R$ 
in table \ref{GeometricalParameters}.  At the same time, we have obtained $N_{\rm e} = n_{0} x_{0}$ in table \ref{BestFitValues} 
and thus $n_{0}$ can be evaluated.  The total mass of the ring, $M_{\rm R}$ 
is also gotten from the following equation with the values of $x_{0}$, $R$ and $n_{0}$ as
\begin{equation}
M_{\rm R} = m_{\rm H} n_{0} 4 \pi^{2} x_{0}^{2} R.
\label{eqn:Calc_MR}
\end{equation}
Results of the above calculations are summarized in table \ref{GaseousProperties}, 
where $\tau_{0}$ is a typical optical depth of the ring tube for Thomson scatterings 
defined as $\tau_{0} = \sigma_{\rm T} n_{0} x_{0}$.

\begin{table}
  \caption{Gaseous properties of the ring matter}
  \begin{center}
    \begin{tabular}{c|ccc}
      \hline
parameter & SMC X-1 & LMC X-4 & Her X-1 \\
     \hline
$T$ & 1.7 $\times 10^{5}$ K & 7.9 $\times 10^{5}$ K & 1.0 $\times 10^{6}$ K \\
$n_{0} x_{0}$ & 2.4 $\times 10^{24}$ cm$^{-2}$ & 3.4 $\times 10^{24}$ cm$^{-2}$ &  3.5 $\times 10^{24}$ cm$^{-2}$ \\
      $\tau_{0}$ & 1.6 & 2.3 & 2.3  \\ 
$n_{0}$ & $5.6 \times 10^{13}$ cm$^{-3}$ & $1.4 \times 10^{14}$ cm$^{-3}$ & $6.6 \times 10^{13}$ cm$^{-3}$  \\ 
$M_{\rm R}$ & $1.4 \times 10^{24}$ g & $5.0 \times 10^{23}$ g & $1.7 \times 10^{24}$ g \\ $t_{C}$ & 1.3 $\times 10^{-2}$ s & 2.3 $\times 10^{-2}$ s & 6.3 $\times 10^{-2}$ s \\
      \hline
    \end{tabular}
  \end{center}
  \label{GaseousProperties}
\end{table}

The matter flowing from the companion star is considered to form the circular ring 
around the compact star in a scheme as discussed in 4.1.
As shown in 4.1.2, the temperature of the ring matter is expected to be $T_{0}$ 
in equation (\ref{eqn:T_0}) when the ring is formed without energy loss.
Temperatures determined from the best fit parameters in table \ref{GaseousProperties} are not largely different from the $T_{0}$ values 
calculated respectively for the three sources within a factor of 3.
On the other hand, 
the typical cooling time-scales of the ring matter, $t_{\rm C}$, are estimated from an equation as
\begin{equation}
t_{\rm C} = \frac{3 k T}{n_{0} \Lambda},
\label{eqn:t_C}
\end{equation}
where $\Lambda$ is a cooling function at the temperature, $T$.
Values of $t_{\rm C}$ for three sources in case of $\Lambda = 10^{-22}$ erg cm$^{3}$ s$^{-1}$ (see e.g. Sutherland and Dopita 1993) are included in table 4.  
We see that those cooling time-scales are much shorter than 
the rotational period of the ring matter, $t_{\phi}$, as
\begin{equation}
t_{\phi} = \frac{2\pi R}{v_{\phi}} = 1.5 \times 10^{4} (\frac{R}{10^{11} \rm{cm}})^{3/2} (\frac{M_{\rm X}}{1.4 M_{\odot}})^{-1/2} s,
\label{eqn:V_phi}
\end{equation}
where, $v_{\phi}$ is the Keplerian circular velocity with the radius $R$.
If the radiative cooling effectively worked, the temperatures must have more largely cooled down to 10$^{4}$ K or so.
This strongly suggests that there should exist some heating on the ring matter 
against the radiative cooling, and that such heating prevent the radiative cooling 
from effectively working in the ring-tube.  
The ring-tube matter is considered to be in a balance between X-ray heating and radiative cooling, as discussed in the next subsection.

\subsection{Accretion properties and presence of X-ray heating}

The observed photon flux in 4 - 20 keV, $F_{0}$, of each of the three sources is given 
as a best fit parameter in table 1.  Assuming the mean photon energy to be 10 keV and 
the distances to SMC X-1, LMC X-4 and Her X-1 are 57 kpc, 49 kpc and 6.1 kpc 
(Leahy \& Abdallah 2014) respectively, their X-ray luminosities, $L_{\rm X}$, are estimated.
The accretion rate, $\dot{M}$, responsible for the luminosity can approximately be evaluated from 
an equation as
\begin{equation}
L_{\rm X} = \dot{M} \frac{G M_{\rm X}}{R_{\rm X}},
\label{eqn:L_X}
\end{equation}
where $R_{\rm X}$ is a radius of the accreted neutron star and is assumed to be 10 km here.
Estimated $L_{\rm X}$ and $\dot{M}$ are given in table \ref{AccretionProperties}.

\begin{table}
  \caption{Accretion properties of the three sources}
  \begin{center}
    \begin{tabular}{c|ccc}
      \hline
parameter & SMC X-1 & LMC X-4 & Her X-1 \\
     \hline
$L_{\rm X}$ & 6.2 $\times 10^{38}$ erg s$^{-1}$ & 1.3 $\times 10^{38}$ erg s$^{-1}$ & 1.1 $\times 10^{38}$ erg s$^{-1}$ \\
$\dot{M}$ & 3.9 $\times 10^{18}$ g s$^{-1}$ & 6.1 $\times 10^{17}$ g s$^{-1}$ &  5.5 $\times 10^{17}$ g s$^{-1}$ \\
$t_{A}$ & 3.6 $\times 10^{5}$ s & 8.2 $\times 10^{5}$ s & 3.0 $\times 10^{6}$ s \\
$\xi$ & 2.5 $\times 10^{2}$ erg cm s$^{-1}$ & 1.1 $\times 10^{2}$ erg cm s$^{-1}$ &  8.5 $\times 10^{1}$ erg cm s$^{-1}$ \\      \hline
    \end{tabular}
  \end{center}
  \label{AccretionProperties}
\end{table}

From these values of $\dot{M}$  and those of the ring mass, $M_{\rm R}$ 
in table 4, we can get the accumulation time of the ring matter, $t_{\rm A}$, 
defined as
\begin{equation}
t_{\rm A} = \frac{M_{\rm R}}{\dot{M}},
\label{eqn:t_A}
\end{equation}
and the values are included in table \ref{AccretionProperties}.
We see that the accumulation time-scales, $t_{\rm A}$, in table \ref{AccretionProperties} are all much 
longer than the respective cooling time-scales, $t_{\rm C}$, in table \ref{GaseousProperties}.
If radiative cooling were really effective, the ring tube must have shrunk in a time-scale as short as $t_{\rm C}$, and $t_{\rm A}$ should have been $\sim t_{\rm C}$.
$t_{\rm A} >> t_{\rm C}$ again suggests that there should be some heating against 
the radiative cooling.
Then, we have calculated typical values of the so-called ionization parameter, $\xi$, 
of the ring by $L_{\rm X} / (n_{0} R^{2}$).
The results are shown in table 5 and we see $\xi \simeq 10^{2}$ erg cm s$^{-1}$ 
for all the three sources.  

\begin{figure}
  \begin{center}
    \FigureFile(80mm,80mm){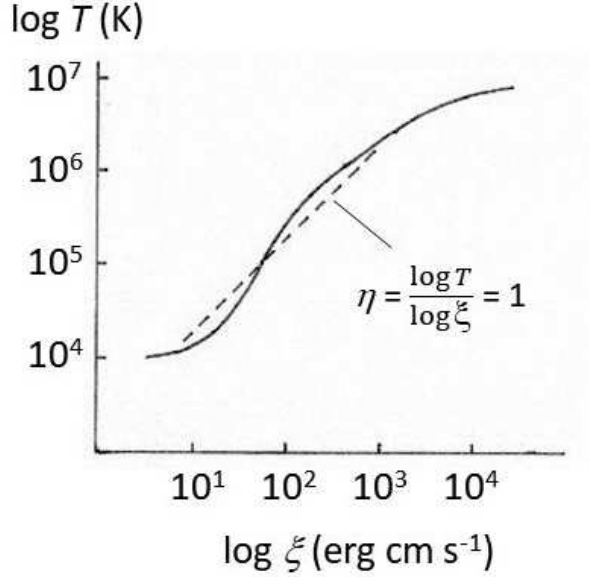}
  \end{center}
  \caption{Schematic diagram on the relation between the ionization parameter, $\xi$ 
and $T$ relation of matter under a balance between radiative cooling and X-ray heating, following Kallman and McCray (1982).}
\label{Xi-T}
\end{figure}

According to theoretical studies on effects of X-ray heating, temperature of matter in a balance between radiative cooling and X-ray heating tends to start 
increasing from $\sim 10^{4}$ K when $\xi$ exceeds $\sim 10$ erg cm s$^{-1}$, 
becomes 10$^{5}$ K around $\xi = 10^{2}$ erg cm s$^{-1}$ and get close to 10$^{6}$ K
before $\xi$ becomes larger than 10$^{3}$ erg cm s${-1}$ (see a schematic diagram 
in figure \ref{Xi-T} which follows figures 7 and 11 in 
Kallman and McCray, 1982).

The $\xi$ and $T$ values estimated from the best fit parameters are 
$\sim 10^{2}$ erg cm s$^{-1}$ and $10^{5} \sim 10^{6}$ K respectively and 
are roughly consistent with the theoretical relation for 
matter in which X-ray heating rate balances with radiative cooling rate.
The gaseous matter in the ring-tube should be in a thermal equilibrium 
under a balance between X-ray heating and radiative cooling.  

\subsection{Excitation of the ring precession under X-ray heating}

Inoue (2012) studied the energetics of the precessing ring on an adiabatic approximation 
as a function of the tilt angle and concluded that a precession of a ring would be excited when the ring matter has sufficiently large thermal energy.

The considerations from present model fits, however, indicate the presence of 
X-ray heating on the ring matter.  Thus, the energetics arguments done by Inoue (2012) cannot directly be applied.
We now have to study whether the precession is possible to be excited or not 
under the presence of X-ray heating.
Results of the study are given in appendix 3, and we see that 
when the parameter, $\eta$ (for the definition, see equation (\ref{eqn:T-xi-eta})) is 
large enough and the temperature is high enough, the ring matter should start precessing.
Even if the ring matter suffers from effect of X-ray heating, the precession would be  possible to be excited.

As seen from figure \ref{Xi-T}, $\eta$ can be larger than unity around 
$\xi \simeq 10^{2}$ erg cm s$^{-1}$.
This is consistent with the $\xi$ values in table 5 and arguments in appendix 3.

Importance of X-ray irradiation has already been pointed out in relation to 
how a precession of an accretion disk is excited, by Pringle (1996), 
Wijers and Pringle (1999), and Oglilvie and Dubus (2001).
Their arguments are, however, on radiation pressure induced by re-emission 
after absorption of X-rays irradiated on a warped thin accretion disk and 
the target situation is different from the present one.
It is discussed in this paper, what happens on matter inflowing from the companion star just before extending to such an accretion disk.

\section{Discussion on accretion flow through the accretion ring}

We have seen above that the precessing accretion ring model well reproduces 
the X-ray light curves folded with the super-orbital periods of the three X-ray 
pulsars and that the physical parameters of the precessing ring obtained from 
the model fits can be reasonably explained.  

\subsection{Steady accretion flow through the accretion ring}

We now discuss how matter from the companion star flows through the accretion 
ring to be finally accreted by the central compact star.

\subsubsection{Ring formation}

Initially, matter coming from the companion star is expected to form a ring 
along a Keplerian circular orbit around the compact star determined 
by a specific angular momentum given with a Coriolis force through the flow 
from the $L_{1}$ point to the ring.
This situation is just what is discussed as the initial circularization 
of the disk formation in Section 4.5 in Frank, King and Raine (2002).  
They obtained an approximate formula for the radius of the ring, $R_{\rm circ}$ as 
\begin{equation}
\frac{R_{\rm circ}}{a} \simeq (1+q) (\frac{b_{1}}{a})^{4}.
\label{eqn:R_circ}
\end{equation}
Here, $b_{1}$ is the distance of the $L_{1}$ point from the compact star and 
is approximately given as
\begin{equation}
\frac{b_{1}}{a} \simeq (0.500 - 0.227 \log q).
\label{eqn:b_1}
\end{equation}
$R_{\rm circ}/a$ are calculated to be $4.7 \times 10^{-2}$, $5.6 \times 10^{-2}$ 
and $1.1 \times 10^{-1}$ for SMC X-1, LMC X-4 and Her X-1 respectively, 
with their respective $q$ values in table \ref{GeometricalParameters}.
These $R_{\rm circ}/a$ values are systematically smaller by a factor of $\sim$ 2 
than the $R/a$ values in table \ref{GeometricalParameters}.
Taking into account that the second term in the right hand side of 
equation (\ref{eqn:R_circ}) is 
the fourth power of the approximate $b_{1}$ value,
we can say that the ring radii estimated on the precessing ring model for the 
super-orbital periodicity are consistent with the simple considerations on the ring 
formation by the matter inflowing through $L_{1}$ to the compact star.

\subsubsection{Thickness of the ring-tube}

The Bernoulli's equation holds for the flow from the $L_{1}$ point to the ring, 
and can be approximated as
\begin{equation}
\frac{1}{2} v^{2} + w - \frac{GM_{\rm X}}{r} = u_{1},
\label{eqn:Bernoulli_Eq}
\end{equation}
at a position with a distance, $r$, from the compact star, 
where $v$, $w$ and $u_{1}$ is a velocity, a specific enthalpy and a total specific energy 
of the inflowing matter.
$u_{1}$ is constant on the assumption of the energy conservation for the flow, 
and is given as a boundary condition at the $L_{1}$ point.
If the inflowing matter forms a ring with a radius, $R$, around the compact star 
with still conserving the total energy, we get
\begin{equation}
\frac{1}{2} v_{\phi}^{2} + w_{\rm R} - \frac{GM_{\rm X}}{R} = u_{1} ,
\label{eqn:Bernoulli_Eq}
\end{equation}
where $v_{\phi}$ and $w_{\rm R}$ are the Keplerian circular velocity and the specific 
enthalpy at $R$.
Since $v_{\phi}^{2} = GM_{\rm X}/R$ and $w = 5 kT_{0}/m_{\rm H}$ for pure hydrogen plasma, the above equation yields the following equation as
\begin{equation}
\frac{5 k T_{0}}{m_{\rm H}} = ( 1 - \varepsilon) \frac{GM_{\rm X}}{2R}.
\label{eqn:T_Rq}
\end{equation}
$T_{0}$ is a temperature of the ring matter when the energy is not lost in the ring formation, 
and we have introduced a parameter, $\varepsilon$, in rewriting $u_{1}$ as
\begin{equation}
u_{1} = - \varepsilon \frac{GM_{\rm X}}{2R}.
\label{eqn:Int_varepsilon}
\end{equation}

On an approximation that a sum of the enthalpy and the kinetic energy at $L_{1}$ is 
a half of the depth of the gravitational potential of the compact star at $L_{1}$, 
$\varepsilon$ should be $R/b_{1}$.
From the arguments on $b_{1}$ and $R$ in 4.1.1., $\varepsilon$ is estimated 
to be $\sim$ 3, and then $T_{0}$ is roughly given as
\begin{equation}
T_{0} \simeq \frac{1}{15} \frac{m_{\rm H}}{ k} \frac{GM_{\rm X}}{R},
\label{eqn:T_0}
\end{equation}
by setting $\varepsilon = 1/3$.
From this equation, $T_{0}$ is calculated to be $5.6 \times 10^{5}$ K, 
$1.5 \times 10^{6}$ K and $1.0 \times 10^{6}$ K respectively for SMC X-1, LMC X-4 
and Her X-1.
$T$ estimated from the best fit parameters in table \ref{GaseousProperties} are 
smaller than above values of $T_{0}$ by a factor of 2 $\sim$ 3 for SMC X-1 and LMC X-4, while $T$ equals to $T_{0}$ for Her X-1.
This can be understood with the following considerations.
Temperature of the inflowing matter should once be as high as 
10$^{6}$ K just after merging with the ring matter.
Although the matter tends to radiatively cool down, X-ray heating prevents it 
from completely cooling and the temperature remains at 
a balance point between X-ray heating and radiative cooling.

As a result, the ring-tube has a fairly large thickness through equation (\ref{eqn:a_def}),
as indicated with $x_{0}/R$ in table \ref{GeometricalParameters}.

\subsubsection{Extension of an accretion disk}

The matter stays in the ring-tube for the accumulation time, $t_{\rm A}$, which is estimated from the model fits to be  $  \sim 10^{6}$ s as seen in table 5.
Then, the matter in the ring should extend to an accretion disk 
and finally be accreted by the central compact star. 
In order for the accretion disk to extend from the ring-tube, the angular momentum 
must be transfered in the ring-tube.

In the so-called $\alpha$ model for the angular momentum transfer in the accretion disk 
(Shakura and Sunyaev 1973, see also Kato et al. 1998; Frank et al. 2002), the in-falling velocity, $v_{\rm r}$, is approximately written 
as 
\begin{equation}
v_{\rm r} = \alpha \Bigl(\frac{h}{r} \Bigr)^{2} v_{\phi},
\label{eqn:alpha_model}
\end{equation}
where $\alpha$ is the viscous parameter, and $h$ and $v_{\phi}$ are the height of the disk, and the Keplerian circular velocity at a disk position with the radius, $r$.
Thus, a viscous spreading time, $t_{\rm V}$, of the ring-tube matter to an accretion disk, 
could approximately be estimated as
\begin{eqnarray}
t_{\rm V} &\simeq& \frac{h}{v_{\rm r}} \nonumber \\
&=& \alpha^{-1}\Bigl (\frac{h}{R}\Bigr)^{-1} \frac{R}{v_{\phi}} \nonumber \\
&=& 1.2 \times 10^{5} \Bigl(\frac{\alpha}{0.1}\Bigr)^{-1} \Bigl(\frac{h/R}{0.2}\Bigr)^{-1}  \Bigl(\frac{R}{10^{11} \rm{cm}}\Bigr)^{3/2} \Bigl(\frac{M_{\rm X}}{1.4 M_{\odot}}\Bigr)^{-1/2} \; \rm{s}, 
\label{eqn:t_V}
\end{eqnarray}
where we have set $r = R$ in equation (\ref{eqn:alpha_model}).
Since the accretion disk extends from the ring tube, the disk height, $h$, should be 
no larger than the ring-tube scale height, $x_{0}$.
From ($x_{0}/R) \simeq 0.2$ in table 1, we get $h/R \lesssim 0.2$.
The viscous parameter, $\alpha$, is currently considered to become  as large as 0.1 when turbulent motions are excited in accretion disks, but it would not 
exceed 0.1.  Thus, we have $\alpha \lesssim 0.1$.
We see $R \simeq 10^{11}$ cm from table \ref{GeometricalParameters}.
By inserting the above three relations into equation (\ref{eqn:t_V}), we obtain 
$t_{\rm V} \gtrsim 1.2 \times 10^{5}$ s.

It would be reasonable that the accumulation time, $t_{\rm A}$ determined 
from the best fit parameters, corresponds to 
the viscous spreading time, $t_{\rm V}$, namely the disk extension time.
In order for $t_{\rm A} = t_{\rm V}$ to hold, we need
\begin{equation}
\Bigl(\frac{\alpha}{0.1}\Bigr) \Bigl(\frac{h/R}{0.2}\Bigr) \simeq 0.1.
\label{eqn:limit_Alpha*h}
\end{equation}

Although the viscous parameter, $\alpha$ is currently considered to be $\sim$ 0.1 
when turbulent motions are excited, it is a result of a kind of order estimation and 
we don't know the exact value.
It could be as small as of the order of 0.01 even in a turbulent situation. 

$h/R$ should also be as small as of the order of 0.01.
As discussed in the next sub-subsection, it is considered that 
shrinkages of relatively dense regions in the ring tube towards the ring-tube center 
could excite turbulent motions in the ring-tube and an accretion disk should extend 
from the central region due to efficient angular momentum transfer induced by 
the turbulent motions.  
The disk should probably be geometrically thin with $h$ much 
smaller than $R$. 

Thus, we can reasonably expect a situation for
equation (\ref{eqn:limit_Alpha*h}) to hold.

\subsubsection{Excitation of turbulent motions in the ring-tube}

Whichever $\alpha$ = 0.1 or 0.01, turbulent motions must be excited in the ring-tube 
for an accretion disk to extend in $\sim 10^{6}$ s.

In the ring tube, the temperature is estimated to be $10^{5} \sim 10^{6}$ K 
and the ionization parameter is $ \sim 10^{2}$ erg cm s$^{-1}$.  
In this parameter range, the logarithmic slope, $\eta$, 
of $T$ against $\xi$ can be larger than unity as discussed in subsection 3.4.
This situation of $\eta > 1$ benefits excitation of the precession, but, at the same time, makes the ring matter thermally unstable.  
A line with $\eta = 1$ in the $\xi$ - $T$ plane 
is a line along which pressure is constant.  
Thus, when $\eta > 1$, a slightly denser region than the average gets denser and denser, and a slightly thiner region gets 
thinner and thinner.
Under this situation, we can expect that a significant part of the ring matter cools down 
even if the X-ray heating rate balances with the radiative cooling rate on average.
Furthermore, the typical Thomson scattering optical depth of the ring tube is as large as or larger than unity as seen in table 4.  Hence, the X-ray heating rate should be reduced 
by the Thomson scattering and radiative cooling should become dominant in the central 
part of the ring tube.
Since the cooled matter begins falling towards the ring tube center, 
there should appear a flow across the rotational motion 
of the ring matter, and turbulent motions should be excited in the central region 
of the ring tube.
Angular momenta of the rotating matter should efficiently be 
transfered through the turbulent motions from the inner side to the outer side of 
the ring tube, 
and then an accretion disk should extend towards the compact star 
while an outflow should arise by receiving the angular momentum transfered from the  inflowing accreted matter.\\

Thus, we can reasonably expect that a steady state is established in the accretion flow 
from the companion star through the accretion ring to the accretion disk extending to the compact star. 

\subsection{Common presence of the accretion ring}

Present study indicates that high mass X-ray binaries would commonly have 
accretion ring around central compact objects.
 
Several low-mass X-ray binaries (LMXBs) are known to repeatedly exhibit absorption dips 
in the X-ray light curves with or close to their respective orbital periods.  
Diaz Trigo et al. (2006) investigated spectral changes during dipping in six bright 
dipping LMXBs in detail.  They found that the spectral changes during dips result 
primarily from an increase in column density and a decrease in the ionization state 
of a highly ionized absorber.  It was also argued that the ionized plasma has a cylindrical 
geometry, since the absorption line properties do not vary strongly with orbital phase 
outside the dips.  This picture is consistent with presence of the X-ray heated accretion 
ring in those dipping sources.  Then, dipping sources are considered to be normal 
LMXBs viewed with large inclination angles and thus such accretion rings should 
be common even to LMXBs.  

From above arguments both for high and low mass X-ray binaries, 
presence of accretion rings shoud be common to binary X-ray sources.

\subsection{Similarities of the broad line regions in AGNs to the accretion rings}

Recently, number of reports on rapid variation of absorption in time scales 
as short as hours in type 2 AGNs (Elvis et al. 2004; Risaliti et al. 2005; Puccetti et al. 2007; Risaliti et al. 2009; Risaliti et al. 2010).  
From such rapid absorption variations, the broad line regions are indicated to be 
locations of the absorbers.
Rapid Compton thick / Compton thin transitions are often observed again from 
type 2 AGNs (Risaliti et al. 2005; Risaliti et al. 2007; Risaliti et al. 2009).  
Variable partial coverings with optically thick X-ray absorbers are suggested to cause 
those rapid transitions.

Spectral changes of type 1 AGNs are also shown to be reproduced by 
the variable partial covering model, but the partially covering matter should be 
warm absorbers in these cases
(Miyakawa et al. 2012; Mizumoto et al. 2014; Iso et al. 2016).
Inoue et al. (2011) discuss that locations of partially covering matter should be 
again in the broad line regions.

In the unified picture of AGNs, the difference between type 1 and 2 AGNs is 
due to a difference in the viewing angle of the central engine from the accretion 
disk plane.
Thus, the line of sight in case of type 1 is considered to go through a relatively outer 
part of the broad line regions, while that of type 2 goes through a region relatively close to 
the equatorial plane. 
From above X-ray observations of type 1 and type 2 AGNs, 
it would be possible to have a picture that there should be clumpy X-ray absorbing  
clouds in the broad line regions and they would get cooler and denser 
as their positions get closer to the equatorial plane.
This seems to be consistent with what are expected to happen in the accretion rings 
as discussed in the previous subsection.
The broad line regions in AGNs could just corresponds to accretion rings at 
the outermost part of accretion disks around the super-massive black holes. 

\section{Summary}

X-ray light curves of three X-ray pulsars, SMC X-1, LMC X-4 and Her X-1, 
folded with their respective super-orbital periods, have been found to be well reproduced 
by the precessing accretion ring model 
that X-rays from a compact object towards us are periodically obscured 
by a precessing accretion ring at the outermost part of an accretion disk 
around the central object. 

From the best fit parameters of the model fit to the super-orbital light curves, 
we see that the optical depth of the accretion ring is commonly 1 $\sim$ 2 
for Compton scattering, the temperature is 10$^{5} \sim 10^{6}$ K, and 
the ionization parameter, $\xi$, is $\sim 10^{2}$ 
erg cm s$^{-1}$ due to X-ray heating from the central X-ray source.  

Simple energetics- and perturbation-arguments indicate that a precession of such a ring is rather stable.  

Such a gray and warm accretion ring inferred from the model fits implies the following evolution scenario on the accretion flow through the accretion ring.

Matter from the companion star carries a certain amount of specific angular momentum and first forms a geometrically thick ring along the Keplerian circular orbit 
around the compact star.
Then, it accumulates there on a viscous spreading time scale, 
during which angular momenta of 
the ring matter are efficiently transfered across the ring-tube and an accretion disk 
extends from the ring-tube.
In circumstances of $T \simeq 10^{5} \sim 10^{6}$ K and $\xi \simeq 10^{2}$ erg cm s$^{-1}$, thermally unstable situations appear in the ring tube and a significant part of 
the ring matter cools down and tends to fall to the ring-tube center.  
Since this cooling flow towards the tube center should cross the steady circulating 
flow in the ring-tube, turbulent motions should be excited in the ring tube.
As a result, angular momenta of the rotating matter should be effectively transferred from the inside to the outside in the ring-tube, 
and an accretion disk should extend from the ring-tube.  

In fact, approximate estimations of the matter accumulation time, $t_{\rm A}$, and 
the viscous spreading time, $t_{\rm V}$, based on the best fit parameters indicate 
that $t_{\rm A} \simeq t_{\rm V}$; namely the viscous spreading time should 
determine the matter accumulation time in the ring-tube.

Finally, we have shown several observational evidences which imply common presences 
of such accretion rings as discussed here in various X-ray sources.

Further study on accretion rings from both theoretical and observational viewpoints are highly desirable.

\bigskip


The author greatly appreciates one of the anonymous referees
for his or her plenty of useful comments.    
This research has made use of the MAXI data provided by RIKEN, JAXA and the MAXI team.

\appendix

\section{Excitation of a precession of an accretion ring}
We assume that all the fluid particles flowing from the companion star 
into the gravitational field of the companion star initially have the same angular momentum per mass, $l$, and form a ring with a radius, $R$, around the compact star.
$R$ is the radius of the Keplerian circular orbit and is given as
\begin{equation}
R = \frac{l^{2}}{GM_{\rm X}},
\label{eqn:R}
\end{equation}
where $G$ is the gravitational constant 
and $M_{\rm X}$ is a mass of the compact star.
Since the ring matter is considered to have thermal energy, 
the ring has a thickness in directions 
perpendicular to the Keplerian circular orbit and forms a tube like structure 
along the circular orbit.  We call this tube structure as the ring tube.   
A simple calculation as in Inoue (2012) shows that residual force 
between the centrifugal force and the gravitational force in a meridian cross section 
in the ring tube can be approximated to act radially towards the tube center, as far as the tube thickness is sufficiently small compared to the ring radius.
Thus, the hydrostatic equation in the tube cross section is approximated as
\begin{equation}
\frac{d P}{d x} = n m_{\rm H} \frac{GM_{\rm X}}{R^{3}} x,
\label{eqn:HydroEq}
\end{equation}
where $x$ is a distance of a position in the tube from the tube center, 
$P$ is a gaseous pressure of the matter in the ring-tube, $n$ is a proton number density 
of the matter, and $m_{\rm H}$ is the proton mass.  
Here, the matter is assumed to be fully 
ionized plasma of pure hydrogen and hence the electron number density is again $n$.
We further approximate the tube matter to be isothermal.
Then, we can solve equation (\ref{eqn:HydroEq}) and get
\begin{equation}
n = n_{0} \exp(-\frac{x^2}{2x_{0}^2}) ,
\label{eqn:n_def}
\end{equation}
where $n_{0}$ is the number density at the center of 
the tube cross section, 
and $x_{0}$ is given as
\begin{equation}
x_{0} = (\frac{2 k T R}{m_{\rm{H}} GM_{\rm{X}}})^{1/2} R,
\label{eqn:a_def}
\end{equation}
where $k$ and $T$ are the Boltzman constant and the temperature of the tube matter.

\begin{figure}
  \begin{center}
    \FigureFile(120mm,80mm){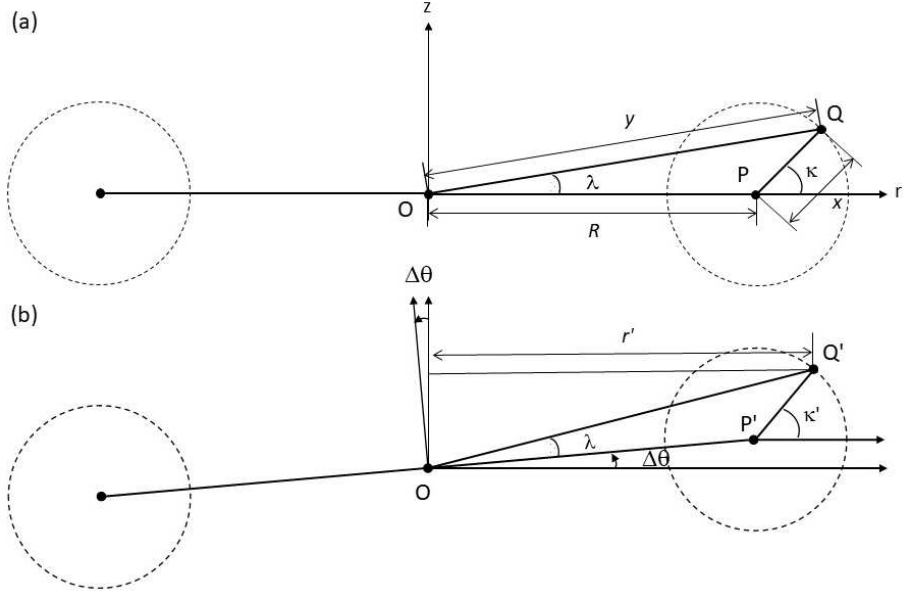}
  \end{center}
  \caption{Configuration change of a meridian cross section of the ring tube by tilting the ring axis with a small angle $\Delta \theta$.  (a) and (b) correspond to figures before and after the change.}\label{Perturbation}
\end{figure}

Inoue (2012) investigated a change of the energetic situation of the ring matter, 
on an adiabatic approximation, as a function of a tilt angle of the ring rotational axis when the ring precesses.  

Here, we add perturbation arguments to the energetics arguments by Inoue (2012).
Let the ring-tube be just on the equatorial plane of the binary system 
and such hydrostatic equilibrium as shown above be established 
in the ring-tube.
Figure \ref{Perturbation}-(a) is a meridian cross section of the ring-tube and we designate the ring center (the position of the compact star) and the ring-tube center with O and P.
We pick up a point, Q, in the cross section and represent its position in two coordinates, 
one in the cylindrical coordinate, ($r$, $z$) around the angular momentum axis of 
the ring matter; the other in the polar coordinate, ($x$, $\kappa$) around the ring-tube 
center.  For the configurations, see figure \ref{Perturbation}-(a), where the distance between Q and O 
and the angle between QO and PO are also introduced with $y$ and $\lambda$.
As seen from figure \ref{Perturbation}-(a), we have the following relations,
\begin{equation}
y = (R^{2} + x^{2} + 2 R x \cos \kappa)^{1/2} \simeq R ( 1 + \frac{x}{R} \cos \kappa ), 
\label{eqn:y} 
\end{equation}
\begin{equation}
\sin \lambda  = \frac{x}{y} \sin \kappa \simeq  \frac{x}{R} \sin \kappa, 
\label{eqn:sin_lambda} 
\end{equation}
\begin{equation}
\cos \lambda  = \frac{R + x \cos \kappa}{y} \simeq 1.
\label{eqn:cos_lambda}
\end{equation}
Here and hereafter, we expand the quantities to the first order of $x/R$ 
since we can approximate $x << R$.

Then, we move the ring-tube by tilting the ring axis by a minute angle, $\Delta \theta$ on the meridian cross section including O, P and Q.
Configurations after the movement is shown in figure \ref{Perturbation}-(b), 
where parameters after it are designated with prime.  
In this movement, we keep the triangle, QPO (and Q'P'O) congruent
and thus  $\lambda$, $x$, $y$ and $R$ do not change.

$r'$ of the position Q' is written as
\begin{equation}
r' = y \cos (\lambda + \Delta \theta) \simeq r [ 1 - \frac{x}{R} (\sin \kappa) \Delta \theta ].
\label{eqn:r^prime}
\end{equation}
Under the assumption of that all the particles in the ring-tube have the same 
angular momentum per mass, the velocity of the particles at Q changes from $v$ to $v'$ by such  a relation as
\begin{equation}
v' = \frac{r}{r'} v.
\label{eqn:v^prime}
\end{equation}
Since the mass flowing rate should be kept at the position Q with the movement, 
the post-number density, $n'$, relates with the pre-number density, $n$, as
\begin{equation}
n' = \frac{v}{v'} n = \frac{r'}{r} n = n [ 1 - \frac{x}{R} (\sin \kappa) \Delta \theta ].
\label{eqn:n^prime}
\end{equation}

Now, we estimate ''kinematic" forces in the r and z directions, 
$F'_{\rm K, r}$ and $F'_{\rm K, z}$, 
induced from the centrifugal force and the gravitational force 
from the compact star, at Q'.
They are
\begin{eqnarray}
F'_{\rm K, r} &=& n' m_{\rm H} [\frac{l^{2}}{r'^{3}} - \frac{GM_{\rm X}}{y^{2}} \cos(\lambda + \Delta \theta)] \nonumber \\
&\simeq& n m_{\rm H} \frac{GM_{\rm X}}{R^{2}} \frac{x}{R} [\cos \kappa + 4  (\sin \kappa) \Delta \theta],
\label{eqn:F'_K,r}
\end{eqnarray}
and
\begin{eqnarray}
F'_{\rm K, z} &=& n' m_{\rm H} \frac{GM_{\rm X}}{y^{2}} \sin (\lambda + \Delta \theta) \nonumber \\
&\simeq& - n m_{\rm H} \frac{GM_{\rm X}}{R^{2}} [\frac{x}{R} \sin \kappa +  ( 1 - \frac{x}{R} \sin \kappa)\Delta \theta ].
\label{eqn:F_K,z^prime}
\end{eqnarray}

On the other hand, ''pressure" forces are given as follows.
Pressure at Q', $P'$, is written as 
\begin{equation}
P' = P (\frac{n'}{n})^{\gamma} \simeq P [1 - \gamma \frac{x}{R} (\sin \kappa) \Delta \theta ],
\label{eqn:P^prime}
\end{equation}
under an adiabatic change approximation.
$\gamma$ is the specific heat ratio.
Thus, the pressure forces in the $x$ and $\kappa$ directions at Q' are
\begin{eqnarray}
F'_{\rm P, x} &=& - \frac{\partial P'}{\partial x} \nonumber \\
&\simeq& - [1 - \gamma \frac{x}{R} (\sin \kappa) \Delta \theta ] \frac{\partial P}{\partial x} + \gamma \frac{P}{R} (\sin \kappa) \Delta \theta ,
\label{eqn:F'_P,x}
\end{eqnarray}
and 
\begin{equation}
F'_{\rm P, \kappa} = -\frac{\partial P'}{x \partial \kappa} \simeq \gamma \frac{P}{R} (\cos \kappa) \Delta \theta,
\label{eqn:F'_P,kappa}
\end{equation}
where we have used a relation as
\begin{equation}
\frac{\partial P}{\partial \kappa} \simeq 0.
\end{equation}
Then, the pressure forces in the $r$ and $z$ directions are calculated as 
\begin{eqnarray}
F'_{\rm P, r} &=& F'_{\rm P, x} \cos \kappa' - F'_{\rm P, \kappa} \sin \kappa' \nonumber \\
&\simeq& - \frac{\partial P}{\partial x} [\cos \kappa - (\sin \kappa) \Delta \theta - \gamma \frac{x}{R}  (\sin \kappa \cos \kappa) \Delta \theta] - \gamma \frac{P}{R} (\sin \kappa \cos \kappa) \Delta \theta,
\label{eqn:F'_P.r}
\end{eqnarray}
and
\begin{eqnarray}
F'_{\rm P, z} &=& F'_{\rm P, x} \sin \kappa' + F'_{\rm P. \kappa} \cos \kappa' 
\nonumber  \\
&=& -\frac{\partial P}{\partial x} [ \sin \kappa +  (\cos \kappa) \Delta \theta] + \gamma \frac{P}{R} \Delta \theta.
\label{eqn:F'_P,z}
\end{eqnarray}
Finally, the total forces in the $r$ and $z$ directions are obtained as
\begin{eqnarray}
F'_{\rm r} &=& F'_{\rm K, r} + F'_{\rm P, r} \nonumber \\
&\simeq& 3 n m_{\rm H} \frac{GM_{\rm X}}{R^{2}} \frac{x}{R} (\sin \kappa) \Delta \theta - \gamma \frac{P}{R} (\sin \kappa \cos \kappa) \Delta \theta, 
\label{eqn:F'_r}
\end{eqnarray}
and
\begin{eqnarray}
F'_{\rm z} &=& F'_{\rm K, z} + F'_{\rm P, z} \nonumber \\
&\simeq& (\gamma \frac{P}{R} - n m_{\rm H} \frac{GM_{\rm X}}{R^{2}}) \Delta \theta + n m_{\rm H} \frac{GM_{\rm H}}{R^{2}} \frac{x}{R} (\cos \kappa + \sin \kappa) \Delta \theta, 
\label{eqn:F'_z}
\end{eqnarray}
with help of equation (\ref{eqn:HydroEq}).

If we integrate $F'_{\rm r}$ and $F'_{\rm z}$ over the tube cross section and divide it 
by the total mass over the cross section, we finally get the average force per mass 
in the two directions as
\begin{equation}
\bar{f'_{\rm r}} \simeq 0,
\label{bar-f'_r}
\end{equation}
and
\begin{eqnarray}
\bar{f'_{\rm z}} &\simeq& (\gamma \frac{P}{R} - \frac{GM_{\rm X}}{R^{2}}) \Delta \theta \nonumber \\
&=& \frac{1}{R} ( \gamma \frac{2kT}{m_{\rm H}} - \frac{GM_{\rm X}}{R}) \Delta \theta.
\label{eqn:bar-f'_z}
\end{eqnarray}
Thus, when
\begin{equation}
 \gamma \frac{2kT}{m_{\rm H}} > \frac{GM_{\rm X}}{R},
\label{eqn:ExctCnd}
\end{equation}
the mattar in the ring-tube receives a further lifting up force, 
and we can say that the ring-tube is unstable against a slight tilting of the rotational axis originally normal to the binary orbital plane.
This is consistent with the result of the energetics arguments by Inoue (2012). 

\section{Error estimation of a folded light curve}

We consider a case that we have obtained $K$ sets of light curves folded into $L$ phase-bins every $N$ cycles with a super-orbital period.
Let $x_{ i,  j}$ be a count rate in the $i$-th phase-bin in the $j$-th light curve set,
and introduce the average count rate in the $i$-th phase-bin, $y_{i}$ over the whole sets of the light curves as
\begin{equation}
y_{i} = \frac{\sum\limits_{j =0}^{K-1} x_{i, j}}{K} .
\label{eqn:Defyi}
\end{equation}
If $x_{i. j}$ randomly fluctuates around the average, $y_{i}$, 
the standard deviation of $x_{i. j}$ around $y_{i}$ can be considered to be 
the error of every $x_{i. j}$ over the $K$ sets.
The purpose of the present analysis is, however, to fit a model light curve to the average light curve, a series of $y_{i}$ from $i$ = 0 to $L$-1.   
Hence, our concern is only on the random fluctuation of $x_{i. j}$ in terms of 
$j$ independent of other phase bins.
Even if a large fluctuation of the amplitude of the light curve exists with keeping 
a constant shape over all the sets, it does not matter on the model fits. 
In order to neglect such a coherent amplitude variation over the $K$ sets of the light curves, we calculate the total count rate over the entire super-orbital period 
for each of the $K$ sets, $z_{j}$ as
\begin{equation}
z_{j} = \sum\limits_{i = 0}^{L - 1} x_{i, j}.
\label{eqn:Defzj}
\end{equation}
Then, we assume that every super-orbital light curve over $K$ sets has the common 
average light curve with a shape with \{ $y_{i}$: $i = 0 \sim L-1$ \} 
but that the amplitude 
of the $j$-th light curve deviates from the average, $y_{j}$ 
with a factor of $z_{j} / \bar{z}$.
$\bar{z}$ is the average of \{ $z_{j}$: $j = 0 \sim K-1$ \}.
An error of the count rate, $x_{i, j}$ in the $i$-th bin of the $j$-th light curve, $\delta x_{,i, j}$ should now be estimated as a deviation from 
its expected count rate $w_{i,j}$ which is defined as
\begin{equation}
w_{i.j} = y_{i} \times \frac{z_{j}}{\bar{z}},
\label{eqn:Defwij}
\end{equation}
and is given by 
\begin{equation}
\delta x_{i,j}^{2} = (x_{i,j} - w_{i,j})^{2}.
\label{eqn:delta_xij}
\end{equation}
Finally, the error of the average count rate in the $i$-th phase-bin, $\delta y_{i}$ is determined by
\begin{equation}
\delta y_{i}^{2} = \frac{\sum\limits_{j=0}^{K-1} \delta x_{i,j}^{2}}{K(K-1)}.
\label{eqn:delta_yij}
\end{equation}

\section{Effect of X-ray heating on the ring temperature}

\subsection{Energetics approach}
Let us assume that the ring mass and the X-ray luminosity are constant while 
the tilt angle changes.
From the ring mass conservation, we have a relation as
\begin{equation}
n_{0} x_{0}^{2} R = \rm constant.
\label{M_R_Const}
\end{equation}
From equation (\ref{eqn:a_def}), we have another relation
\begin{equation}
x_{0} \propto T^{1/2} R^{3/2}.
\label{eqn:a-T-R}
\end{equation}
From above two relations, we get
\begin{equation}
T \propto n_{0}^{-1} R^{-4}.
\label{eqn:T-n0-R}
\end{equation}

Here, we introduce a relation between $T$ and $\xi$ with a logarithmic slope of $\eta$ as 
\begin{equation}
T \propto \xi^{\eta}.
\label{eqn:T-xi-eta}
\end{equation}
Since $\xi = L_{\rm X} / (n_{0} R^{2})$ and $L_{\rm X}$ is constant,
\begin{equation}
T \propto (n_{0} R^{2})^{-\eta}.
\label{eqn:T-xi}
\end{equation}
From equations (\ref{eqn:T-n0-R}) and (\ref{eqn:T-xi}), we obtain
\begin{equation}
T \propto R^{2\eta/(1 - \eta)}.
\label{eqn:T-R-eta}
\end{equation}
According to Inoue (2012), $R$ increases as the tilt angle of the ring axis increases.
Then, in case that $\eta > 1$, $T$ decreases  
and the thermal energy decreases as the tilt angle increases.
Hence, when the thermal energy is sufficiently large, the amount of energy decrease 
of the thermal energy associate with the tilt angle increase can be larger than that of energy increase of the rotational and 
gravitational energy and the energy minimum should appear at a certain, non-zero tilt angle.

\subsection{Perturbation approach}
Under the isothermal approximation, we assume that 
the temperature, $T$, has initially been at a balance point between the heat input rate 
through the X-ray heating and the heat loss rate through the radiative cooling, 
with integrated over the cross section of the ring-tube.
We approximate that $T$ corresponds to that given by the averaged ionization 
parameter over the ring, $\bar{\xi}$.  The over-line means averaging over the relevant region here and hereafter.

Then, let us move the ring as figure \ref{Perturbation} in appendix 1 and $\Delta P$ be 
a change of the pressure after the movement, 
which is written as
\begin{equation}
\Delta P = ( \frac{\Delta n}{n} + \frac{\Delta T}{T}) P.
\label{eqn:DPparP}
\end{equation} 
Since $L_{\rm X}$ is considered to be constant and distances of every positions 
in the ring-tube cross section to the central X-ray source do not vary with the movement, the change of $\xi$ is governed only by a $n$ change.
Thus, we approximate 
\begin{eqnarray}
\frac{\Delta T}{T} &=& \overline{\frac{1}{T} \frac{dT}{d \xi} \frac{\partial \xi}{\partial n}  \Delta n}
\nonumber \\
&=& - \eta \frac{\overline{\Delta n}}{\bar{n}},
\label{eqn:DTperT_1}
\end{eqnarray}
where $dT/d\xi$ should be calculated from a theoretical relation between $T$ and $\xi$ 
at the initial temperature, and $\eta$ is the logarithmic slope there (defined 
in  equation  (\ref{eqn:T-xi-eta})).  $\partial \xi/\partial n$ should be from a relation of 
$\xi = L/(n y^{2})$.
Since
$\Delta n/n$ is given from equation (\ref{eqn:n^prime}) as
\begin{equation}
\frac{\Delta n}{n} = - \frac{x}{R} (\sin \kappa) \Delta \theta,
\label{eqn:Dnpern}
\end{equation}
we can obtain $\Delta T/T$ by averaging $\sin \kappa$ over the relevant region.
Although the average of $\sin \kappa$ over the entire cross section is zero, 
averaging is now done for each of two regions, the upper half ( $0 \le \kappa \le \pi$) 
and the lower half ($\pi \le \kappa \le 2\pi$) of the ring-tube cross section,  
in order to clarify the effect of X-ray heating on the ring-tube due to the movement.
The result is
\begin{equation}
\frac{\Delta T}{T} = \pm \eta (\frac{2}{\pi})^{1/2} \frac{x_{0}}{R} \Delta \theta,
\label{eqn:DTperT_2}
\end{equation}
where + and - corresponds to the case of 
the upper half and the lower half of the ring-tube cross 
section, respectively.

From equations (\ref{eqn:DPparP}), (\ref{eqn:Dnpern}) and (\ref{eqn:DTperT_2}), 
the pressure gradient forces in the $x$ and $\kappa$ directions induced by the ring-tube movement are
\begin{eqnarray}
\Delta F_{\rm P, x} &=& - \frac{\partial \Delta P}{\partial x} \nonumber \\
&=& - \frac{\partial P}{\partial x} [-\frac{x}{R} \sin \kappa \pm \eta (\frac{2}{\pi})^{1/2} \frac{x_{0}}{R} ] \Delta \theta + \frac{P}{R} (\sin \kappa) \Delta \theta \label{eqn:DF_x} \\
\Delta F_{\rm P, \kappa} &=& - \frac{\partial \Delta P}{x \partial \kappa} \nonumber \\
&=& \frac{P}{R} (\cos \kappa) \Delta \theta \label{eqn:DF_kappa}.
\end{eqnarray}
Above two equations give us the pressure gradient force induced in the $z$ direction as
\begin{equation}
\Delta F_{\rm P, z} = - \frac{\partial P}{\partial x} [-\frac{x}{R} \sin^{2} \kappa \pm \eta (\frac{2}{\pi})^{1/2} \frac{x_{0}}{R} \sin \kappa ] \Delta \theta + \frac{P}{R} \Delta \theta,
\label{eqn:DF_Pz}
\end{equation}
while we see from equation (\ref{eqn:F_K,z^prime})
\begin{equation}
\Delta F_{\rm K, z} = - n m_{\rm H} \frac{GM_{\rm X}}{R^{2}} [\frac{x}{R} \sin \kappa +  ( 1 - \frac{x}{R} \sin \kappa) \Delta \theta ].
\label{eqn:DF_Kz}
\end{equation}
By adding the above two equations and using equation (\ref{eqn:HydroEq}), 
the total induced force in the $z$ direction is
\begin{equation}
\Delta F_{\rm z} = n m_{\rm H} \frac{GM_{\rm X}}{R^{2}} [\eta (\frac{2}{\pi})^{1/2} \frac{x_{0}}{R} \frac{x}{R} |\sin \kappa| + \frac{x_{0}^{2}}{R^{2}} + \frac{x}{R} (\sin \kappa) -1] \Delta \theta.
\label{eqn:DF_z}
\end{equation}
By integrating the above equation over the tube cross section and dividing it 
by the total mass over the cross section, we finally get the average force per mass as
\begin{eqnarray}
f_{\rm z} &=& \frac{GM_{\rm X}}{R^{2}} [(\eta \frac{2}{\pi}  + 1) \frac{x_{0}^{2}}{R^{2}} -1] \Delta \theta \nonumber \\
&=& [(\eta \frac{2}{\pi} +1) \frac{2kT}{m_{\rm H}} - \frac{GM_{\rm X}}{R^{2}}] \Delta \theta.
\label{eqn:f_z_XH}
\end{eqnarray}
Thus, when 
\begin{equation}
(\eta \frac{2}{\pi} +1) \frac{2kT}{m_{\rm H}} > \frac{GM_{\rm X}}{R^{2}},
\label{eqn:UnstaCnd}
\end{equation}
the ring tube should be lifted up further than the initial movement.
This is consistent with the result in the previous subsection.




\begin{thebibliography}{}
\bibitem[Clarkson et al.(2003a)]{key-1}
  Clarkson, W.I., Charles, P.A., Coe, M.J., Laycock, S., Tout, M.D., \& Wilson, C.A., \ 2003a, MNRAS, 339, 447
\bibitem[Clarkson et al.(2003)]{key-1}
  Clarkson, W.I., Charles, P.A., Coe, M.J., \& Laycock, S.\ 2003b, MNRAS, 343, 1213
\bibitem[Diaz Trigo et al.(2006)]{key-1}
  Diaz Trigo, M., Parmar, A.N., Boirin, L., Mendez, M., \& Kaastra, J.S.\ 2006, A\&A, 445, 179
\bibitem[Eggleton (1983)]{key-1}
  Eggleton, P.P.\ 1983, ApJ, 268, 368
\bibitem[Elvis et al.(2004)]{key-1}
  Elvis, M., Risaliti, G., Nicastro, F., Miller, J.M., Fiore, F., \& Puccetti, S.\ 2004, ApJL, 615, L25
\bibitem[Falanga et al.(2015)]{key-1}
  Falanga, M., Bozzo, E., Lutovinov, A., Bonnet-Bidaud, J.M., Fetisova, Y., \& Puls, J.\ 2015, A\&A, 577, A130
\bibitem[Frank et al.(2002)]{key-1}
  Frank., J., King, A. \& Raine, D.\ 2002, Accretion Power in Astrophysics (Cambridge, Cambridge University Press)
\bibitem[Gerend et al.(1976)]{key-1}
  Gerend, D., \& Boynton, P.E.\ 1976, ApJ, 209, 562
\bibitem[Giacconi et al.(1973)]{key-1}
  Giacconi, R., Gursky, H., Kellogg, E., Levinson, R., Schreier, E., \& Tananbaum, H.\ 1973, ApJ, 184, 227
\bibitem[Gruber et al.(1984)]{key-1}
  Gruber, D.E., \& Rothschild, R.E.\ 1984, ApJ, 283, 546
\bibitem[Hu et al.(2011)]{key-1}
  Hu, C., Chou, Y., Wu, M.,Yang, T., \& Su, Y.\ 2011, ApJ, 740, 67
\bibitem[Heemskerk et al.(1989)]{key-1}
  Heemskerk, M.H.M., \& van Paradijs, J.\ 1989, A\&A, 223, 154
\bibitem[Inoue et al.(2011)]{key-1}
  Inoue, H., Miyakawa, T., \& Ebisawa, K.\ 2011, PASJ, 63, S669
\bibitem[Inoue(2012)]{key-1}
  Inoue, H.\ 2012, PASJ, 64, 40
\bibitem[Iso et al.(2016)]{key-1}
  Iso, N., Ebisawa, K., Sameshima, H., Mizumoto, M., Miyakawa, T., Inoue, H., \& Yamasaki, H.\ 2016, PASJ, 68, S27
\bibitem[Jones et al.(1976)]{key-1}
  Jones, C., \& Forman, W.\ 1976, ApJL, 209, L131
\bibitem[Kallman et al.(1982)]{key-1}
  Kallman, T.R., \& McCray, R.\ 1982, ApJS, 50, 263
\bibitem[Kato et al.(1998)]{key-1}
  Kato, S., Fukue, J. \& Mineshige, S.\ 1998, Black Hole Accretion Disks (Kyoto, Kyoto University Press)
\bibitem[Katz (1973)]{key-1}
  Katz, J.I.\ 1973, NaturePhysSci, 246, 87
\bibitem[Kotze et al.(2012)]{key-1}
  Kotze, M.M., \& Charles, P.A.\ 2012, MNRAS, 420, 1575
\bibitem[Lang et al.(1981)]{key-1}
  Lang, F.L., et al.\ 1981, ApJL, 246, L21
\bibitem[Leahy et al.(2014)]{key-1}
  Leahy, D.A., \& Abdallah, M.H.\ 2014, ApJ, 793, 79
\bibitem[Levine et al.(1982)]{key-1}
  Levine, A.M., \& Jernigan, J.G.\ 1982, ApJ, 262, 294
\bibitem[Matsuoka et al.(2009)]{key-1}
  Matsuoka, M., et al.\ 2009, PASJ, 61, 999
\bibitem[Miyakawa et al.(2012)]{key-1}
  Miyakawa, T., Ebisawa, K., \& Inoue, H.\ 2012, PASJ, 64, 140
\bibitem[Mizumoto et al.(2014)]{key-1}
  Mizumoto, M., Ebisawa, K., \& Sameshima, H.\ 2014, PASJ, 66, 122
\bibitem[Neilsen et al.(2009)]{key-1}
  Neilsen,J., Lee, J.C., Nowak, M.A., Dennerl, K., \& Vrtilek, S.D.\ 2009, ApJ, 696, 182
\bibitem[Ogilvie et al.(2001)]{key-1}
  Ogilvie, G.I., \& Dubus, G.\ 2001, MNRAS, 320, 485
\bibitem[Parmar et al.(1985)]{key-1}
  Parmar, A.N., Pietsch, W., McKechnie, S., White, N.E., Truemper, J., Voges, W., \& Barr, P.\ 1985, Nature, 313, 119
\bibitem[Priedhorsky \& Holt(1987)]{key-1}
  Priedhorsky, W.C., \& Holt, S.S.\ 1987, SSRv, 45, 291
\bibitem[Pringle (1996)]{key-1}
  Pringle, J.E.\ 1996, MNRAS, 281, 357
\bibitem[Puccetti et al.(2007)]{key-1}
  Puccetti, S., Fiore, F., Risaliti, G., Capalbi, M., Elvis, M., \& Nicastro, F.\ 2007, MNRAS, 377, 607
\bibitem[Risaliti et al.(2005)]{key-1}
  Risaliti, G., Elvis, M., Fabbiano, G., Baldi, A., \& Zezas, A.\ 2005, ApJL, 623, L93
\bibitem[Risaliti et al.(2007)]{key-1}
  Risaliti, G., Elvis, M., Fabbiano, G., Baldi, A., Zezas, A., \& Salvati, M.\ 2007, ApJL, 659, L111
\bibitem[Risaliti et al.(2009a)]{key-1}
  Risaliti, G., et al.\ 2009a, MNRAS, 393, L1
\bibitem[Risaliti et al.(2009b)]{key-1}
  Risaliti, G., et al.\ 2009b, ApJ, 696, 160
\bibitem[Risaliti et al.(2010)]{key-1}
  Risaliti, G., Elvis, M., Bianchi, S., \& Matt, G.\ 2010, MNRAS, 406, L20
\bibitem[Shakura et al.(1973)]{key-1}
  Shakura, N.I., \& Sunyaev, R.A.\ 1973, A\&A, 24, 337
\bibitem[Sutherland et al.(1993)]{key-1}
  Sutherland, R.S., \& Dopita, M.A.\ 1993, ApJS, 88, 253
\bibitem[Wen et al.(2006)]{key-1}
  Wen, L., Levine, A.M., Corbet, R.H.D., \& Bradt, H.V.\ 2006, ApJS, 163, 372
\bibitem[Wijers et al.(1999)]{key-1}
  Wijers, R.A.M.J., \& Pringle, J.E.\ 1999, MNRAS, 308, 207
\bibitem[Wojdowski et al.(1998)]{key-1}
  Wojdowski, P., Clark, G.W., Levine, A.M., Woo, J.W., \& Zhang, S.N.\ 1998, ApJ, 502, 253


\end{thebibliography}
\end{document}